\documentclass[conference,letterpaper]{IEEEtran}
\IEEEoverridecommandlockouts
\usepackage[letterpaper, top=1.92cm, bottom=3.2cm, left=1.92cm, right=1.92cm]{geometry}
\usepackage{pifont}
\usepackage{tabularx}
\usepackage{graphicx}
\usepackage{fancyhdr}
\usepackage{amsmath,amssymb,amstext,amsfonts}
\usepackage{siunitx}
\usepackage{diagbox}
\usepackage{bm}
\usepackage{algorithm}
\usepackage{array}
\usepackage{cite}
\usepackage{epstopdf}
\usepackage{balance}
\usepackage{algorithmicx}
\usepackage{algpseudocode}
\usepackage{epsfig}
\usepackage{braket}
\usepackage{tabulary}
\usepackage{diagbox}
\usepackage{amsthm}
\usepackage{color}
\usepackage{verbatim}
\usepackage{multirow} 
\usepackage{hyperref}
\usepackage{booktabs}
\usepackage{rotating}
\usepackage{adjustbox}
\usepackage{tabularx}
\usepackage{subfigure}

\newif\iftracked
\trackedfalse
\iftracked
    \newcommand{\revisioncolor}{blue}
\else
    \newcommand{\revisioncolor}{black}
\fi

\makeatletter
\IEEEtriggercmd{\reset@font\normalfont\footnotesize}
\makeatother
\IEEEtriggeratref{1}

\begin{document}
\bstctlcite{BSTcontrol}

\title{Electrodermal Activity (EDA) for Stress Detection: A Comprehensive Survey and Benchmark}


\author{Gang~Liu, Zixiong~Ning
\thanks{Gang Liu and Zixiong Ning are with the Department of Electrical Engineering, University of Notre Dame, Notre Dame, IN 46556 USA (e-mail: gangliu2004@outlook.com; zning2@nd.edu).}

\thanks{\textcolor{\revisioncolor}{The code will be made publicly available upon acceptance.}}
}

\maketitle
\label{loc_code_availability}

\begin{abstract}
\label{loc_abstract}
\color{\revisioncolor}
Electrodermal activity (EDA) is widely used for pervasive stress monitoring because of its sensitivity to sympathetic arousal and its computational efficiency. However, research remains fragmented: studies concentrate on a small subset of available datasets, preprocessing and evaluation protocols are inconsistent, and widely adopted methodological conventions remain unvalidated. Existing reviews summarize these practices without evaluating them under a common protocol. To bridge these gaps, this paper presents a comprehensive survey and benchmark for EDA-based stress detection. We review the technical pipeline from preprocessing to modeling and compile 26 public datasets and 22 representative unimodal studies to inform future dataset selection and methodological design. Under a subject-independent protocol spanning five diverse datasets, we benchmark decomposition, filtering, and channel-input choices alongside a broad range of machine learning (ML) and deep learning (DL) models in both in-domain and cross-domain transfer settings. Our results provide the first multi-dataset validation of cvxEDA, the default decomposition method, show that feature-based ML outperforms end-to-end (E2E) DL, and identify ecological validity as a key factor in cross-domain transfer. We conclude with directions toward robust, data-efficient, and trustworthy deployment.
\color{black}
\end{abstract}


\section{Introduction}
\label{sec_introduction}
Stress is a growing health burden, compounded by information overload, fast-paced lifestyles, and blurred work--life boundaries \cite{impactofremotework_Intro_2022,WHO_Intro_2025}. Persistent ``distress'' (as opposed to performance-enhancing ``eustress'') sustains activation of the sympathetic nervous system (SNS) \cite{mentalhealth_Intro_2024,2KindofStress_Intro_1946}, elevating cortisol levels and increasing the risk of cardiovascular disease, immune suppression, and cognitive decline \cite{vchronicstressonhealth_Intro_2015}. 
Questionnaires carry self-report bias, and biomarkers such as cortisol require obtrusive sampling and delayed laboratory analysis; neither supports continuous real-time monitoring \cite{PSS_Table_Intro_2023,Cortisol_Intro_2023}.
Research has therefore turned to physiological signals such as electrocardiogram (ECG) \cite{HR_Intro_2018}, electroencephalogram (EEG) \cite{EEG_Intro_2025}, and photoplethysmogram (PPG) \cite{PPG_Intro_2022}. These signals track stress-induced physiological changes. Automated detection built on them has become the mainstream approach to intervening before distress escalates \cite{MentalStressDetection_RelatedSurvey_2021,WearablesDevice4Stress_Intro_2023}. Multimodal fusion pushes accuracy higher, but the improvement is marginal relative to the added computation and the burden of wearing multiple sensors \cite{MultiModal_Intro_2025}.

In contrast, unimodal electrodermal activity (EDA) supports pervasive stress monitoring with a single low-cost sensor and modest computational requirements \cite{ZhuLiLi_JBHI_PureEDA_ML_2023}. EDA measures sweat gland activity directly regulated by the SNS, making it sensitive to sympathetic arousal without parasympathetic interference \cite{EDA_Textbook_2012}. 
Under standard protocols, EDA has shown correlations of 0.7--0.9 with stress levels; it alone has also been reported to match or outperform its fusion with ECG and PPG, and comparisons across physiological modalities rank it among the most predictive individual signals for stress \cite{ZhuLiLi_JBHI_PureEDA_ML_2023,EDAisBest_PureEDA_ML_2022,DeepLearningApproach_PureEDA_ML_2025,EDAisBest_ML_MultiModal_2022,EnsembleClassificationModel_DL_MultiModal_2021}. Its compatibility with portable sensors supports continuous monitoring and edge inference in driving, workplace, educational, and human--computer interaction settings \cite{Driving1_Application_2025,Safety_Application_2019,Safety2_Application_2024,Work1_Application_2017,Study1_Application_2023,Study2_Application_2019,Study3_Application_2022,Study5_Application_2023,Study6_Application_2024,HCI1_Application_2019,HCI2_Application_2021,HCI3_Application_2024,CDACVX_DecompoistionComparision_2022}. Together, these properties give EDA a practical balance among predictive performance, hardware cost, and computational demand.

{\color{\revisioncolor}
This advantage has attracted a large body of work, and several practices have settled into conventions: acquisition has standardized on the Empatica E4 at 4~Hz, and cvxEDA has become the default decomposition method. Yet these defaults propagated by repetition rather than by validation. Three critical gaps still remain.

First, existing research evaluates on too few datasets. Most public EDA datasets are never used, so cross-domain generalization goes unverified.
Second, the pipelines are neither standardized nor validated. Filtering, segmentation, and feature-engineering choices differ across studies, and some studies still leak test information into fitted quantities, so reported accuracies cannot be compared. Even the decomposition step, now dominated by cvxEDA, has been compared only within single datasets, and those comparisons disagree.
Third, existing surveys remain descriptive. EDA-specific reviews describe the methodological inconsistencies but do not test them; broader stress-detection surveys focus on multimodal fusion and say little about the preprocessing and decomposition choices in unimodal EDA. None catalogs the available datasets alongside the studies that use them, and none benchmarks the pipeline under a common protocol.
This paper addresses these gaps through the following contributions:
}
\begin{itemize}
    \item \textbf{Methodology Integration \& Resource Compilation:} We present the first survey to combine a review of the unimodal EDA pipeline with a benchmark. The review covers artifact removal, decomposition, and modeling. It also catalogs 26 public datasets and 22 representative unimodal studies to foster community reproducibility.
    \item \textbf{Cross-Domain Benchmark:} We establish a leave-one-subject-out (LOSO) protocol across five datasets to benchmark EDA decomposition, filtering, and input choices alongside a broad range of ML and DL models. We further test cross-domain generalization by holding out each dataset in turn and training on one, two, or all four remaining datasets, with and without fine-tuning (FT) on the target.
    \item \textbf{Future Outlook:}  We outline directions for closing the gap between benchmark performance and real-world use: adapting to domain shift and individual variability, reducing reliance on labeled data through generative methods and self-supervised learning (SSL), and making models efficient, private, and physiologically interpretable on wearables.
\end{itemize}

The remainder of this paper is organized as follows: Section~\ref{sec_background_and_related_work} covers EDA fundamentals, existing surveys, and public datasets. Section~\ref{sec_technical_review_of_eda} reviews the technical pipeline from artifact removal to modeling. Section~\ref{sec_experiments} reports the benchmark. Section~\ref{sec_future_outlook_and_discussion} discusses future directions. Finally, Section~\ref{sec_conclusion} concludes the paper.
\label{sec_introduction_end}
\section{Background and Related Work}
\label{sec_background_and_related_work}
To identify EDA-based stress-detection studies, we searched IEEE Xplore, PubMed, Web of Science, and Google Scholar using combinations of the keywords ``electrodermal activity'', ``skin conductance'', ``Galvanic Skin Response (GSR)'', ``stress detection'', ``stress recognition'', ``mental workload'', and ``wearable''. We focused on human-subject studies published between 2010 and 2026 that (i) recorded EDA signals, (ii) included explicit stress labels, and (iii) reported quantitative performance of a computational model. 
\subsection{EDA Signal Fundamentals}
\label{sec_eda_signal_fundamentals}
\begin{figure}[h]
    \centering
    \setlength{\subfigtopskip}{0pt}
    \setlength{\subfigbottomskip}{0pt}
    \subfigure[EDA Response]{%
        \includegraphics[width=0.49\columnwidth]{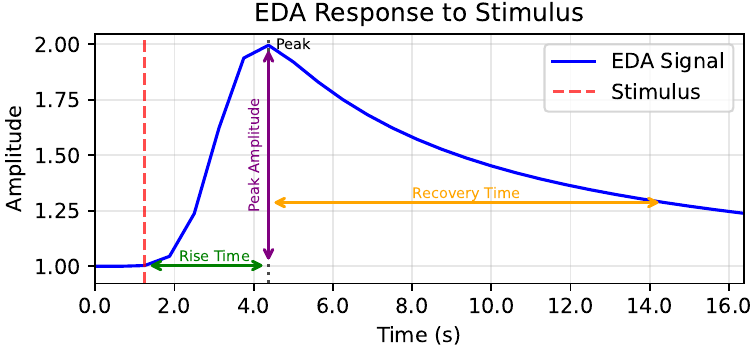}\label{fig_eda_response}}%
    \hfill
    \subfigure[EDA Decomposition]{%
        \includegraphics[width=0.49\columnwidth]{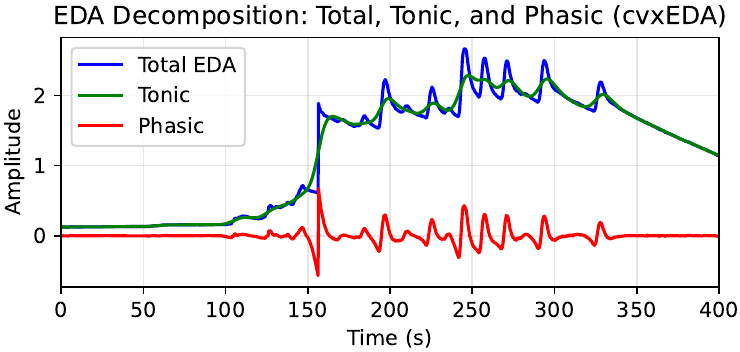}\label{fig_eda_decomposition}}
    \vspace{0pt}                
    \caption{Illustration of an EDA signal.}
    \label{fig_eda_signal}
    \vspace{-8pt}                
\end{figure}
\begin{sidewaystable*}
\centering
\footnotesize
\renewcommand{\arraystretch}{1}
\begin{adjustbox}{max width=\textheight, max totalheight=0.86\textwidth, center}
\begin{tabular}{
p{0.08\textwidth}
p{0.03\textwidth}
p{0.10\textwidth}
p{0.12\textwidth}
p{0.31\textwidth}
p{0.31\textwidth}
}
\toprule
\textbf{Ref./Year} & \textbf{\# Refs.} & \textbf{Signal Modality} & \textbf{Core Focus} & \textbf{Main Contributions} & \textbf{Main Limitations}
\\
\midrule
\multicolumn{6}{l}{\textit{\textbf{Part I: EDA-Specific Surveys}}} \\
\midrule
1.\cite{PureEDAStressDetection_RelatedSurvey_2019}/2019& 14 & EDA Only & EDA for Stress and Arousal-Related States & 1. An early attempt to collect EDA-based stress studies, predating dedicated EDA reviews. & 1. Limited literature coverage (only 14 articles).\newline 2. Outdated (published in 2019) and missing recent advances.\newline 3. Primarily a descriptive list of studies, lacking in-depth insight or analysis.\\
2.\cite{CollectionandSignalProcessing_RelatedSurvey_2020}/2020& 89 & EDA Only
& EDA Data Acquisition and Signal Processing 
& 1. Systematically reviews EDA data acquisition and signal processing techniques.
\newline 2. Summarizes recording devices, electrode technology, and spectral methods.
\newline 3. Discusses data quality assessment and artifact processing methods.
& 1. Focuses on the signal processing front-end, without covering model construction.
\newline 2. Lacks a direct comparison of different methods and guidance on their selection (e.g., better decomposition algorithms).\\
3.\cite{ML4ST_preprint_RelatedSurvey_2020}/2020& 103 & EDA Only & EDA with ML for Stress Detection & 1. Systematic review of the full pipeline for EDA-based stress detection.\newline 2. Summarizes and compares the performance of existing features and models.\newline 3. Emphasizes the balanced advantage of EDA in terms of accuracy and computational cost.
& 1. The methodological summary (e.g., datasets and decomposition algorithms) lacks sufficient depth.\newline 2. Conflates research on emotion recognition with stress detection.\newline 3. Has not undergone peer review (preprint). \\
4.\cite{DetectingEmotionsLearningContexts_RelatedSurvey_2021}/2021& 86 & EDA Only & EDA-based Emotion Recognition in Learning Contexts & 1. Outlines the process of EDA-based emotion detection in educational contexts and points out the lack of standardization.\newline 2. Systematically summarizes contradictory findings in studies linking EDA and learning. 
& 1. The application scenario is limited to education, reducing generalizability.\newline 2. The methodological discussion is relatively basic and provides an incomplete comparison of models and features.
\\
5.\cite{EDAstressincare_RelatedSurvey_2025}/2025& 74 & EDA Only & EDA-based Stress Sensing in Care Environments & 1. Bridges theory and practice, emphasizing the potential of EDA in special care settings (e.g., dementia).\newline 2. Provides a detailed statistical analysis of subjects, devices, and other elements in the reviewed studies.
& 1. The reviewed studies focus mostly on healthy individuals in laboratory settings, with insufficient coverage of real care contexts.\newline 2. The analysis is based primarily on simple statistics and lacks methodological synthesis and insight.
\\
6.\cite{RecentTrends_EDA_EmotionRecognition_RelatedSurvey_2026}/2026& 99 & EDA Only & EDA Signal Processing and DL for Emotion Recognition &
1. Reviews EDA decomposition algorithms and time-, frequency-, and time-frequency-domain feature families.
\newline   2. Compiles publicly available EDA datasets for emotion recognition with acquisition-protocol details.
\newline 3. Summarizes EDA-specific analysis methods and organizes them into four categories: nonlinear, symbolic, topological, and graph-based.
& 1. Promises a comparative evaluation using key performance metrics but rates methods as High/Medium/Low based on the literature.
\newline 2. Identifies artifacts as core obstacles for EDA, but does not review artifact detection or removal methods.
\\
\midrule

\multicolumn{6}{l}{\textit{\textbf{Part II: Multimodal Surveys (Physiological \& General)}}} \\
\midrule
7.\cite{MentalStressDetection_RelatedSurvey_2021}/2021& 75 & Multimodal Physiological Signals & Mental Stress Detection using Wearable Devices & 1. Classifies studies by sensor type and application environment, summarizes the correlation between physiological signals and stress, and highlights the importance of EDA.\newline 2. Reviews the strengths, limitations, and future directions of the field.& 1. Does not systematically summarize public datasets.\newline 2. Provides an overly descriptive analysis that lacks quantitative comparisons and in-depth methodological discussion.
\\
8.\cite{Generalizablemachinelearning_RelatedSurvey_2023}/2023& 87 & Multimodal Physiological Signals & Generalizability of ML for Stress Detection & 1. Follows a standard ML pipeline to systematically review commonly used datasets and pipelines.\newline 2. Explicitly identifies and emphasizes the poor generalizability of models that rely on single, small datasets. & 1. The summary of datasets and methods remains mainly descriptive and lacks deep insights.\newline 2. The analysis of the generalization problem remains superficial and does not explore solutions in depth.
\\
9.\cite{KnowledgeWorkEnvironments_RelatedSurvey_2025}/2025& 202 & Multimodal Data (Physiological, Behavioral, Environmental) & Continuous Stress Monitoring in Knowledge Work Environments & 1. Reviews the integrated use of physiological, behavioral, and environmental data in knowledge work scenarios.\newline 2. Emphasizes the importance of personalized models and user acceptance. & 1. The scope is limited to knowledge work and lacks cross-domain comparisons.\newline 2. Discussions of topics such as user acceptance are overly theoretical and lack support from empirical data.
\\
10.\cite{DL_RelatedSurvey_2025}/2025& 185 & Multimodal Data (Physiological, Speech, Facial, Social Media) & Application of DL in Multimodal Stress Detection & 1. Provides a comprehensive overview of DL applications in multimodal stress detection, highlighting the advantage of automatic feature extraction.\newline 2. Details the methods, performance, and public datasets for each modality. & 1. Focuses exclusively on DL, neglecting traditional ML approaches.\newline 2. Provides relatively limited discussion and depth regarding EDA.
\\
\midrule
\multicolumn{6}{l}{\textit{\textbf{Proposed Work}}} \\
\midrule
\textbf{This Study} & 220 & Primarily EDA & A Comprehensive Survey and Benchmark & 1. \textbf{Integrated Survey and Resources}: Presents the first survey to combine a review of the unimodal EDA pipeline with a benchmark, cataloging 26 public datasets and 22 studies while identifying the protocol inconsistencies that prevent direct comparison.\newline 2. \textbf{Benchmark}: Provides the first multi-dataset validation of cvxEDA under a five-dataset LOSO benchmark and systematically compares preprocessing, models, and transfer, showing that consistent protocols, effective representations, and ecologically valid data matter more than model complexity or simple data aggregation.\newline 3. \textbf{Field-Level Insight and Research Agenda}: These findings define priorities for addressing domain shift, label scarcity, and real-world constraints through domain adaptation, data-efficient learning, privacy, interpretability, and efficient wearable deployment. & 1. Does not evaluate advanced learning paradigms such as SSL and domain adaptation.  \newline  2. Does not incorporate recent AI technologies such as LLMs or agentic systems. \\
\bottomrule
\end{tabular}
\end{adjustbox}
\caption{Comparison of related surveys. The table is divided into two sections: \textbf{Part I} reviews surveys explicitly targeting EDA, while \textbf{Part II} covers high-impact surveys of broader multimodal stress detection. The \textbf{`\# Refs.'} column indicates the total number of references cited in each survey.}
\label{tab_comparison_of_related_surveys}
\end{sidewaystable*}

EDA, also known as GSR, measures changes in skin conductance caused by the sympathetic regulation of sweat gland activity. It is commonly recorded from the wrist, fingertips, or palms and is used as an indicator of physiological arousal associated with stress responses. As shown in Fig.~\ref{fig_eda_response}, EDA is nonstationary, with mean skin conductance levels typically ranging from 2--20~$\mu$S. A typical skin conductance response has a rise time of approximately 1--3~s and a half-recovery time of 2--10~s \cite{FeatureCombine_EDA_EmotionRecognition_2021}. As illustrated in Fig.~\ref{fig_eda_decomposition}, EDA can be decomposed into tonic (0--0.05~Hz) and phasic (0.05--0.5~Hz) components. The tonic component, or Skin Conductance Level, represents slow baseline fluctuations and is influenced by environmental conditions and individual baselines. The phasic component, or Skin Conductance Response (SCR), consists of transient responses to sympathetic activation. Overlapping SCRs create a complex signal waveform, while features such as amplitude and response rate indicate phasic sympathetic arousal. 
\subsection{Comparison with Other Surveys}
\label{sec_comparison_with_other_surveys}
As summarized in Table~\ref{tab_comparison_of_related_surveys}, we review existing surveys closely related to this study. To ensure the quality and relevance of the comparison, we selected representative surveys from two main categories: those explicitly targeting EDA as the primary research subject and high-impact surveys focusing on the broader field of stress detection. 

\subsubsection{EDA-Centric Surveys}\label{sec_eda_centric_surveys} Early studies, such as \cite{CollectionandSignalProcessing_RelatedSurvey_2020}, pioneered research on EDA signal acquisition and processing techniques but did not extend to downstream modeling and detection pipelines. An early survey was the first to explicitly target EDA-based stress detection~\cite{PureEDAStressDetection_RelatedSurvey_2019}; it covered a limited number of papers and primarily listed existing studies without providing methodological insights or analyzing recent advances. Subsequent studies began to cover the technical pipeline but were often constrained by narrow application scenarios, such as emotion recognition exclusively in educational contexts \cite{DetectingEmotionsLearningContexts_RelatedSurvey_2021} or specific care environments \cite{EDAstressincare_RelatedSurvey_2025}, limiting their generalizability. One preprint review conflates the distinct fields of ``stress detection'' and ``emotion recognition''~\cite{ML4ST_preprint_RelatedSurvey_2020}.
A recent review of unimodal EDA-based emotion recognition covers EDA decomposition algorithms and relevant emotion datasets~\cite{RecentTrends_EDA_EmotionRecognition_RelatedSurvey_2026}. 

\subsubsection{Surveys Centered on Multimodal Stress Detection}\label{sec_surveys_centered_on_multimodal_stress_detection} Compared with unimodal EDA studies, research in this category is generally more mature and systematic. These surveys typically review the field from the perspectives of signal input, feature extraction, and classification models (covering both feature-based ML and E2E DL) \cite{Pinge_StressWearablesDifferentDevice_relatedsurvey_2024,SmartGadgetsStress_relatedsurvey_2025,StressReview_relatedsurvey_2025,predictingstressepisodes_relatedsurvey_2025}.
Some studies extend their scope to specialized modalities, such as biochemical sensors (e.g., salivary cortisol) \cite{predictingstressepisodes_relatedsurvey_2025}, contactless technologies such as spectral imaging and rPPG \cite{ContactlessrPPG_relatedsurvey_2025}, and even behavioral, environmental (e.g., CO$_2$ concentration), or social media data \cite{KnowledgeWorkEnvironments_RelatedSurvey_2025, DL_RelatedSurvey_2025, SmartGadgetsStress_relatedsurvey_2025}. Specific scenarios, such as knowledge work environments \cite{KnowledgeWorkEnvironments_RelatedSurvey_2025} and student populations \cite{StudentStress_Application_relatedsurvey_2024}, have also been discussed. These surveys address issues shared across modalities, such as the ``lab-to-wild'' gap, device heterogeneity, and overfitting on small datasets, but breadth comes at the cost of depth. Few examine the details specific to EDA, such as preprocessing parameters, artifact removal, and the comparison of decomposition algorithms \cite{MentalStressDetection_RelatedSurvey_2021, Generalizablemachinelearning_RelatedSurvey_2023}. They emphasize the importance of EDA \cite{SmartGadgetsStress_relatedsurvey_2025, predictingstressepisodes_relatedsurvey_2025} without supplying the pipeline-level detail needed to act on it.

\subsection{Datasets}
\label{sec_datasets}
As shown in Table~\ref{tab_public_eda_stress_datasets}, we compiled information on 26 public datasets containing EDA signals relevant to stress detection. Although previous surveys have discussed datasets in affective computing \cite{Cognitive_datasetSurvey_2020,9_datasetSurvey_2025,58_datasetSurvey_2025,32_datasetSurvey_2025}, they often do not fully meet the specific needs of EDA-based stress detection research. For instance, \cite{Cognitive_datasetSurvey_2020} focuses primarily on cognitive load and does not cover the broader context of psychological stress; \cite{9_datasetSurvey_2025} is limited in scope, covering only 9 datasets, of which just 5 involve EDA; and although \cite{58_datasetSurvey_2025,32_datasetSurvey_2025} summarize numerous stress datasets, they do not specifically target scenarios in which EDA is the core modality.

\begin{sidewaystable*}
\centering
\small
\renewcommand{\arraystretch}{1.25}
\begin{adjustbox}{max width=\textheight, max totalheight=\textwidth, center}
\begin{tabular}{
  p{0.14\textwidth}
  p{0.04\textwidth}
  p{0.05\textwidth}
  p{0.07\textwidth}
  p{0.18\textwidth}
  p{0.18\textwidth}
  p{0.28\textwidth}
  p{0.04\textwidth}}
\toprule
\textbf{Dataset Name} & \textbf{Year} & \textbf{Subj.} & \textbf{Dur.} & \textbf{Physiological Signals} & \textbf{EDA Device/$f_s$} & \textbf{Description} & \textbf{Link} \\
\midrule
1. RWDriving \cite{RealWolrdDriving_Dataset_2005} & 2005 & 9 (17 records) & 65--93~min & ECG, EDA, EMG, RESP & FlexComp/31~Hz & Real-world driving & \href{https://physionet.org/content/drivedb/1.0.0/}{Link} \\

2. SWELL \cite{SWELL_Dataset_2014} & 2014 & 25 & 3~hrs & ECG, EDA, Voice & TMSI Mobi/2048~Hz & Knowledge work, email interruptions & \href{https://cs.ru.nl/~skoldijk/SWELL-KW/Dataset.html}{Link} \\

3. Neuro \cite{NonEEGNeurological_Dataset_2017} & 2016 & 20 & 37~min & ACCEL, EDA, HR, SpO2, ST & Affectiva Q/8~Hz & Physical, cognitive, and emotional stressors & \href{https://physionet.org/content/noneeg/1.0.0/}{Link} \\

4. WESAD \cite{WESAD_Dataset_2018} & 2018 & 15 & 2~hrs & ACCEL, BVP, ECG, EDA, EMG, HR, RESP, ST & RespiBAN/700~Hz, Empatica E4/4~Hz & Baseline, stress, amusement, and meditation states & \href{https://ubi29.informatik.uni-siegen.de/usi/data_wesad.html}{Link} \\

5. AffectiveROAD \cite{AffectiveROAD_Road_2018} & 2018 & 10 (13 records) & 2~hrs & ACCEL, BVP, EDA, HR, IBI, ST & Empatica E4/4~Hz & Real-world driving & \href{https://www.media.mit.edu/tools/affectiveroad/}{Link} \\

6. CLAS \cite{CLAS_Dataset_2019} & 2019 & 62 & 30~min & ACCEL, ECG, EDA, PPG & Shimmer GSR Unit/256~Hz & Cognitive tasks and affective visual stimuli & \href{https://ieee-dataport.org/open-access/database-cognitive-load-affect-and-stress-recognition}{Link} \\

7. Toadstool \cite{Toadstool_Dataset_2020} & 2020 & 10 & 1~hr & ACCEL, BVP, EDA, HR, IBI, ST & Empatica E4/4~Hz & Playing the game ``Super Mario Bros.'' & \href{https://datasets.simula.no/toadstool/}{Link} \\

8. MAUS \cite{MAUS_Dataset_2021} & 2021 & 22 & 47~min & ECG, EDA, PPG (Finger/Wrist) & Procomp Infiniti/256~Hz & N-back cognitive task & \href{https://ieee-dataport.org/open-access/maus-dataset-mental-workload-assessment-n-back-task-using-wearable-sensor}{Link} \\

9. VerBIO \cite{VerBIO_Dataset_2022} & 2022 & 55 & Scenario-Dependent & ACCEL, BVP, ECG, EDA, ST & Empatica E4/4~Hz & Public speaking anxiety & \href{https://hubbs.engr.tamu.edu/resources/verbio-dataset/}{Link} \\

10. EmpathicSchool \cite{EmpathicSchool_Dataset_2022} & 2022 & 20 & 1.3~hrs & ACCEL, BVP, EDA, HR, IBI, ST & Empatica E4/4~Hz & Completing various tasks in front of a computer & \href{https://zenodo.org/records/7097178}{Link} \\

11. Wearable Exam Stress \cite{Exam_Dataset_2022} & 2022 & 10 & 6~hrs & ACCEL, BVP, EDA, HR, IBI, ST & Empatica E4/4~Hz & Real exam & \href{https://physionet.org/content/wearable-exam-stress/1.0.0/}{Link} \\

12. Nurse Stress Prediction \cite{Nurse_Dataset_2022} & 2022 & 15 & 1~week & ACCEL, BVP, EDA, HR, IBI, ST & Empatica E4/4~Hz & Hospital work & \href{https://datadryad.org/dataset/doi:10.5061/dryad.5hqbzkh6f}{Link} \\

13. StressID \cite{StressID_dataset_2023} & 2023 & 65 & 35~min & ECG, EDA, RESP & BioSignalsPlux System/500~Hz & 11 tasks including cognitive, emotional, and speech & \href{https://project.inria.fr/stressid/}{Link} \\

14. UBFC-Phys \cite{UBFCPhys_dataset_2023} & 2023 & 56 & 9~min & BVP, EDA (derived: rPPG, PRV) & Empatica E4/4~Hz & Speech and mental arithmetic tasks & \href{https://sites.google.com/view/ybenezeth/ubfc-phys}{Link} \\

15. BIOSTRESS \cite{BIOSTRESS_dataset_2023} & 2023 & 20 & 15 or 60+ min & ACCEL, BVP, EDA, HR, IBI, ST & Empatica E4/4~Hz & Stressful interview and final exam & \href{https://data.mendeley.com/datasets/z9y2fn38fp/1}{Link} \\

16. K-EmoPhone \cite{KEmoPhone_dataset_2023} & 2023 & 77 & 7~days & ACCEL, EDA, HR, RRI, ST & Microsoft Band 2/5~Hz & Seven days of daily life & \href{https://zenodo.org/records/7606611}{Link} \\

17. ForDigitStress \cite{ForDigitStress_dataset_2023} & 2023 & 40 & 85~min & EDA, PPG, Voice & IOM Biofeedback Sensor/NA & Job interview scenario & \href{https://hcai.eu/fordigitstress/#gain-access}{Link} \\

18. MUMTG \cite{MUMTG_dataset_2024} & 2024 & 20 & NA & BVP, EDA, HR, IBI, ST & Empatica E4/4~Hz & Building blocks task & \href{https://data.mendeley.com/datasets/b7c2h6cbc6/1}{Link} \\

19. EmpaticaE4Stress \cite{EmpaticaE4Stress_dataset_2024} & 2024 & 29 & 40~min & ACCEL, BVP, EDA, HR, ST & Empatica E4/4~Hz & Simulated work & \href{https://data.mendeley.com/datasets/kb42z77m2g/2}{Link} \\

20. WorkStress3D \cite{workstress_dataset_2024} & 2024 & 20 & 8.75~hrs & ACCEL, BVP, EDA, ST & Empatica E4/4~Hz & In-office work day & \href{https://data.mendeley.com/datasets/t93xcwm75r/5}{Link} \\

21. EmoPairCompete \cite{EmoPairCompete_dataset_2024} & 2024 & 28 & 72~min & ACCEL, BVP, EDA, HR, ST & Empatica E4/4~Hz & Competitive Tangram puzzle task & \href{https://zenodo.org/records/11151714}{Link} \\

22. CIRVR \cite{CIRVR_dataset_2024} & 2024 & 15 & 20~min & ACCEL, BVP, EDA, HR, IBI, ST & Empatica E4/4~Hz & Simulated virtual job interview & \href{https://lab.vanderbilt.edu/rasl/dataset-and-or-code-request/}{Link} \\

23. Structured-Exercise \cite{InducedStressandStructuredExercise_dataset_2025} & 2025 & 36 & Protocol-Dependent & ACCEL, BVP, EDA, HR, IBI, ST & Empatica E4/4~Hz & Induced stress and structured exercise & \href{https://physionet.org/content/wearable-device-dataset/1.0.1/}{Link} \\

24. Mindset \cite{MindsetMatters_Personalization_2026} & 2026 & 23 & 1~hr & ACCEL, BVP, ECG, EDA, HR, IBI, RRI, ST & Empatica E4/4~Hz & Singing and Stroop color-word test & \href{https://social-dynamics.net/stress/mindset/}{Link} \\

25. CAN-STRESS \cite{CANSTRESS_dataset_2026} & 2026 & 82 & 24~hrs & ACCEL, BVP, EDA, HR, IBI, ST & Empatica E4/4~Hz & Daily life & \href{https://zenodo.org/records/14842061}{Link} \\

26. Stress-Resilience \cite{HCIStress_dataset_2026} & 2026 & 35 & 31~min & ACCEL, BVP, EDA, EEG, fNIRS, HR, IBI, ST & Empatica E4/4~Hz & Resource-management and communication task & \href{https://physionet.org/content/neuro-stress-resilience-hci/1.0.0/}{Link} \\
\bottomrule
\end{tabular}
\end{adjustbox}
\caption{Summary of 26 publicly available datasets for stress detection featuring EDA. \textbf{Abbreviations}: Accelerometer (ACCEL), Blood Volume Pulse (BVP), Electrocardiogram (ECG), Electrodermal Activity (EDA), Electroencephalogram (EEG), Electromyogram (EMG), functional Near-Infrared Spectroscopy (fNIRS), Heart Rate (HR), Inter-Beat Interval (IBI), Photoplethysmogram (PPG), Pulse Rate Variability (PRV), Respiration (RESP), Remote Photoplethysmography (rPPG), R-R Interval (RRI), Blood Oxygen Saturation (SpO2), Skin Temperature (ST). Duration refers to the recording time per subject. }
\label{tab_public_eda_stress_datasets}
\end{sidewaystable*}
We catalog the key attributes of each dataset (subject count, experiment duration, signal modalities, collection devices/frequencies, experimental paradigms, and access links), providing researchers with a reference guide. However, despite the public availability of these datasets, their practical use presents several challenges:
\begin{itemize}
    \item \textbf{Data accessibility barriers:} Some datasets (e.g., VerBIO \cite{VerBIO_Dataset_2022}, EmpathicSchool \cite{EmpathicSchool_Dataset_2022}, StressID \cite{StressID_dataset_2023}, K-EmoPhone \cite{KEmoPhone_dataset_2023}, ForDigitStress \cite{ForDigitStress_dataset_2023}, CIRVR \cite{CIRVR_dataset_2024}, and Mindset \cite{MindsetMatters_Personalization_2026}) are not available for direct download and require researchers to submit application forms or contact the authors by email.
    
    \item \textbf{Technical usability barriers:} Early datasets (e.g., RWDriving \cite{RealWolrdDriving_Dataset_2005}) are often limited by outdated acquisition devices and protocols, while others (e.g., SWELL \cite{SWELL_Dataset_2014}) use proprietary formats (e.g., S00) that require specific software for parsing.
    
    \item \textbf{Uncertain release status:} The release status of several promising datasets remains uncertain. Highly valuable resources, such as LAUREATE \cite{LAUREATE_dataset_2023} (longitudinal ecological study), MMSD \cite{MMSD_Dataset_2022} (multimodal with cortisol), and MuSE \cite{MuSE_Dataset_2020} (real exam stressors), have been described in the literature with the intention of being made open source. Yet, at the time of this survey, they remain unreleased.
\end{itemize}
Two clear trends emerge from Table~\ref{tab_public_eda_stress_datasets}. First, acquisition hardware has become increasingly standardized. The three datasets released before 2018 used different devices sampling at 31, 2048, and 8~Hz, whereas 18 of the 23 datasets released since then include Empatica E4 recordings at 4~Hz. This consistency makes a shared protocol practical and motivates us to resample every dataset to 4~Hz (Section~\ref{sec_experimental_setup}). However, it also means that most evidence comes from one vendor's wrist-worn sensor, so it remains unclear whether the results apply to other devices \cite{ReproducibilityConsumerWearableSensors_FutureDirection_2025}. Second, recording duration has increased substantially. Datasets released before 2022 were mostly limited to laboratory sessions of at most three hours, whereas Nurse \cite{Nurse_Dataset_2022}, K-EmoPhone \cite{KEmoPhone_dataset_2023}, and CAN-STRESS \cite{CANSTRESS_dataset_2026} contain continuous recordings spanning one day to one week. This shift extends coverage from isolated stress episodes to everyday monitoring.
\label{sec_background_and_related_work_end}
\section{Technical Review of EDA}
\label{sec_technical_review_of_eda}
\subsection{Artifact Removal}
\label{sec_artifact_removal}
EDA signals are susceptible to artifacts from multiple sources. Strategies to mitigate them fall into two categories: \textit{signal correction} (suppression) and \textit{artifact identification} (detection and rejection).

\subsubsection{Signal Correction}\label{sec_signal_correction} Traditional non-adaptive methods, such as exponential smoothing and a low-pass filter (LPF), often fail to handle high-amplitude artifacts and may distort valid signals \cite{ExponentialSmoothing_ArtifactRemoval_2011,LowPassFilter_ArtifactRemoval_2014}. To address this issue, adaptive denoising frameworks have been developed. Chen et al. introduced a stationary wavelet transform approach, but it proved sensitive to parameter tuning \cite{StationaryWaveletTransform_ArtifactRemoval_2015}. Subsequent improvements include modeling wavelet coefficients with a Laplace distribution to replace a computationally expensive Gaussian mixture model \cite{Unsupervisedmotionartifactdetectionn_ArtifactRemoval_2018} and using biophysical models with extended Kalman filters for robust noise suppression \cite{ArtifactandNoisesuppression_ArtifactRemoval_2015}. In specific contexts such as driving, dual-hand fusion strategies have successfully filtered sharp spikes caused by steering maneuvers \cite{ECGEDA_ML_MultiModal_2019,Twohands_ArtifactRemoval_2022}. More recent methods employ DL tools, such as convolutional autoencoders that reconstruct clean EDA \cite{ConvolutionalAutoencoder_ArtifactRemoval_2022}. Subsequent work has distilled denoising models into lightweight versions that retain reconstruction fidelity during resource-constrained wearable deployment \cite{MemoryEfficient_ArtifactRemoval_2026}. A different perspective treats temperature as a potential confounder of EDA and attempts to remove its influence~\cite{RemoveTempInfluence_ArtifactRemoval_2026}, although the reported results did not support this hypothesis.

\subsubsection{Artifact Identification}\label{sec_artifact_identification} Early supervised methods used SVMs to classify artifacts but relied heavily on manual annotation \cite{TaylorAutomatic_ArtifactRemoval_2015}. Zhang et al. argued that supervised approaches scale poorly and demonstrated the effectiveness of unsupervised learning for this task \cite{Unsupervisedmotionartifactdetectionn_ArtifactRemoval_2017}. To avoid the ``black-box'' nature of ML, Kleckner et al. proposed a transparent, rule-based method that uses heuristic physiological constraints, thereby enhancing the algorithm's interpretability \cite{AutomatedQualityAssessment_ArtifactRemoval_2018}.
\subsection{EDA Decomposition}\label{sec_eda_decomposition}
Broadly, EDA decomposition splits the raw signal into tonic, phasic, and noise components and uses the phasic component to infer underlying neural activity. Bach et al. established an early psychophysiological modeling framework. They modeled event-related responses using linear convolution models that were validated for linearity and time invariance and refined through optimized response functions and filtering \cite{Bach_TSA_Decomposition_2009,Bach_ModellingEvent_Decomposition_2010,Bach_ImprovedAlgorithm_Decomposition_2013}. They described spontaneous activity using the skin-conductance integral, dynamic causal modeling (DCM) of sudomotor bursts, and matching pursuit for computational efficiency \cite{Bach_QuantificationArousal_Decomposition_2010,Bach_DCM_spontaneous_Decomposition_2011,Bach_matchingpursuitalgorithm_Decomposition_2015}. To separate the tonic and phasic components, Benedek et al. proposed Continuous and Discrete Decomposition Analysis (CDA and DDA), both of which model the SCR shape with a biexponential Bateman function but recover the phasic driver as a continuous series and as discrete non-negative impulses, respectively \cite{CDA_Decomposition_2010,DDA_Decomposition_2010}. A parallel line of research frames decomposition as sparse recovery, ranging from overcomplete physiological dictionaries \cite{SparseRep_Decomposition_2015} to the widely adopted convex and noise-robust cvxEDA \cite{cvxEDA_Decomposition_2016}. Later methods largely refine these directions and include compressed sensing \cite{CompressedSensing_Decomposition_2017}, non-negative sparse deconvolution (sparsEDA) \cite{sparsEDA_Decomposition_2018}, block coordinate descent \cite{GeneralizedCrossValidationBasedBlock_Decomposition_2019}, continuous-time system identification \cite{SparseDeconvolution_Decomposition_2019}, state-space modeling (BayesianEDA) \cite{BayesianEDA_Decomposition_2022}, multichannel decomposition \cite{SparseMultichannelDecomposition_Decompoistion_2023}, a data-driven dynamic model (UDM) \cite{UnifiedDynamicModel_Decompoistion_2025}, and orthogonal subspace projection (ospEDA) \cite{ospEDA_Decompoistion_2026}.

Early tools were mostly integrated toolboxes: Ledalab (including CDA and DDA) and PsPM (which incorporates the former SCRalyze and a large collection of analysis algorithms developed by Bach et al.). Later algorithms, such as cvxEDA, sparsEDA, BayesianEDA, UDM, and ospEDA, are distributed directly through GitHub, improving their accessibility and extensibility.\footnote{Tool links: \href{http://www.ledalab.de/}{Ledalab}, \href{https://bachlab.github.io/PsPM/}{PsPM}, \href{https://github.com/lciti/cvxEDA}{cvxEDA}, \href{https://github.com/fhernandogallego/sparsEDA}{sparsEDA}, \href{https://github.com/computational-medicine-lab/BayesianEDA-For-Sharing}{BayesianEDA}, \href{https://github.com/huisophiewang/UDM_EDA}{UDM}, \href{https://github.com/yongbin98/ospEDA}{ospEDA}.} General physiological signal-processing libraries, such as NeuroKit2 \cite{neurokit_toolbox_2021} and BioSPPy \cite{BioSPPy_toolbox_2024}, have also integrated EDA-processing algorithms, further lowering the technical barrier for researchers.

Comparative studies suggest that decomposition performance is task-dependent \cite{IndicesofPain_DecompoistionComparision_2021}. For stimulus-locked responses, SCRalyze showed greater sensitivity than Ledalab \cite{2_DecompoistionComparision_2014}. For stress detection, \cite{sparsLedaCVX_DecompoistionComparision_2021} found sparsEDA was computationally fastest and slightly more accurate than Ledalab and cvxEDA on a single dataset, while CDA and cvxEDA were comparable \cite{CDACVX_DecompoistionComparision_2022}. For emotion recognition, CDA, DCM, cvxEDA, sparsEDA, and BayesianEDA differed by less than two percentage points under the same classifier and task settings \cite{6_DecompoistionComparision_2024}.
{\color{\revisioncolor}Prior stress-specific comparisons were limited to single datasets and identified no consistent winner. Section~\ref{sec_preprocessing_and_input_comparison} provides the first multi-dataset evaluation of the cvxEDA convention.}

{\color{\revisioncolor}
\begin{sidewaystable*}
\centering
\small
\renewcommand{\arraystretch}{1.3}
\begin{adjustbox}{max width=\textheight, max totalheight=0.76\textwidth, center}
\begin{tabular}{
  p{0.05\textwidth}
  p{0.035\textwidth}
  p{0.18\textwidth}
  p{0.25\textwidth}
  p{0.25\textwidth}
  p{0.15\textwidth}
  p{0.12\textwidth}
  p{0.18\textwidth}}
\toprule
\textbf{Ref.} & \textbf{Year} & \textbf{Dataset(s)} & \textbf{Preprocessing} & \textbf{Feature Information\newline(Domain) (Selection, Count)} & \textbf{Model(s)} & \textbf{Validation} & \textbf{Reported Performance} \\
\midrule

\multicolumn{8}{l}{\textit{\textbf{Part I: Machine Learning (ML) Approaches}}} \\
\midrule
1.\cite{automaticworkloadestimation_PureEDA_ML_2016}& 2016 & Self-collected; 5 mental arithmetic tasks & 50~Hz notch, min-max norm, 500~ms moving avg, downsampling to 50~Hz, 3.5~s window & 55 (Time, Freq, Time-Freq, Quefrency) (PCA, 90\% variance) & \textbf{Combined SVM}, PNN & Nested 5-fold CV per subject$^{\dagger}$ & 90.0\%/80.2\% Acc (3/5-class) \\
2.\cite{physicalandpsychologicalstressorsn_PureEDA_ML_2017}& 2017 & Self-collected; running \& TSST & 5~Hz LPF, 20~s-window-detrend $\to$ 3 streams, 3~min window & 33 (Time) (Wrapper method, 5) & \textbf{LDA}, QDA, SVM, KNN & LOOCV$^{\dagger}$ & 95.1\% Acc (2-class stress type classification) \\
3.\cite{RealDriving_PureEDA_ML_2018}& 2018 & RWDriving & LPF, 5~min window & 18 (Time) (Fisher projection, 2) & LDA & Leave-one-sample-out$^{\dagger}$ & 81.82\% Acc (3-class stress level) \\
4.\cite{XGBOOST_PureEDA_ML_2019}& 2019 & WESAD & 4~Hz downsample, 2~Hz LPF, 60~s window (30~s overlap), 0.2~Hz decomp.& 195 (Time, Freq, Entropy, Wavelet) (Pearson $|r|>0.9$ then importance-Top-9) & \textbf{Extreme Gradient Boosting (XGB)}, DT, RF, AdaBoost, LDA, KNN & LOSO & 92.38\% (chest)/89.92\% (wrist) F1 (2-class) \\
5.\cite{PRESURGERY_PureEDA_ML_2020}& 2020 & Self-collected; pre-surgery patients & 5~Hz LPF, detrend, discard first 15~mins, 5~s window, KNN artifact detection & 57 (Time, Wavelet) (Stat. test + wrapper, 4-7) & Localized Supervised Learning & 30/11 subject split$^{\ddagger}$ & 85.06\% Acc (3-class stress level) \\
6.\cite{OpenSourceTool_PureEDA_ML_2020}& 2020 & WESAD& 20~Hz downsample, 1~s Moving Average Filter, Min-Max Norm& No specific number (Statistical features, CNN-embedding features) (None)& \textbf{KNN}, SVM, RF, NB & 10-fold CV$^{\dagger}$& 91.6\% Acc for DL features, 90 \% for statistical features (2-class) \\
7.\cite{DSVM_PureEDA_ML_2020}& 2020 & Self-collected; calm/stress movies & 4~Hz LPF, gaussian smoothing, CDA (Ledalab) decomp., 1--40~s window & 23 (Time, Morphological, Statistical, Freq) (None) & \textbf{Deep SVM (DSVM)}, various traditional classifiers & 70/15/15 split$^{\dagger}$ & 92\% F1 (2-class) \\
8.\cite{electronicnose_PureEDA_ML_2020}& 2021 & Self-collected; academic stress & 5~min sample, z-score norm & Raw signal input (LDA for dimensionality reduction) & \textbf{SVM}, LDA, KNN & 5-fold CV$^{\dagger\ddagger}$ & 100\% Acc (GSR), 96\% Acc (e-nose) (2-class) \\
9.\cite{CDACVX_DecompoistionComparision_2022}& 2022 & Self-collected; Stroop test & Downsample, smooth, z-score norm, CDA/cvxEDA decomp. & 4 (Time) (None) & ELM & 6-fold CV$^{\dagger}$ & 95.56\% (CDA)/94.45\% (cvxEDA) Acc (3-class) \\
10.\cite{FeasibilityStudy_PureEDA_ML_2022}& 2022 & WESAD, VerBIO & 30~s window (no overlap), cvxEDA decomp. & 7 (Time) (None) & KNN, LR, RF & 90/10 subject split & 85.3\% (VerBIO)/85.7\% (WESAD) Acc (2-class) \\
11.\cite{EDAisBest_PureEDA_ML_2022}& 2022 & WESAD, CLAS & 30~s window (no overlap), cvxEDA decomp. & 7 (Time) (None) & \textbf{SEL}, SVM, KNN, RF, NB, LR & Not specified & 86.4\% (WESAD)/72.3\% (CLAS) Acc (2-class) \\
12.\cite{LowCost_PureEDA_ML_2022}& 2022 & WESAD & 8~Hz upsample, 5~Hz LPF, 60~s window, 0.25~s stride, cvxEDA decomp. & 87 (Time) (Forward selection, 5) & \textbf{AdaBoost}, RF, SVM & LOSO$^{\ddagger}$ & 97.03\% Acc (2-class) \\

13.\cite{AcuteStressStateClassification_PureEDA_ML_2023}& 2023 & Self-collected; TSST & Artifact Removal, cvxEDA decomp. & 11 (Time, Freq) (SVM-RFE, 8 for 2-class, 6 for 4-class) & \textbf{SVM-RFE}, RF, LDA & LOSO $^{\ddagger}$& 94.62\%/91.03\% bal.(2-class)/75.00\% (4-class) Acc \\
14.\cite{FrequencySpectrumAnalysis_PureEDA_ML_2023}& 2023 & WESAD & 12~s window (5~s overlap) & $3(3m+1)$ (Freq) (None) & DT, 1/10-NN, RF, SVM, Bagged SVM, AdaBoost & LOSO & 84.87\% (2-class)/69.89\% (3-class) Acc \\
15.\cite{ZhuLiLi_JBHI_PureEDA_ML_2023}& 2023 & CLAS, Neuro, VerBIO, WESAD & 30~s window (no overlap), cvxEDA decomp. & 7 (Time) (None) & KNN, SVM, NB, LR, RF & 90/10 subject split & 92.9\% (VerBIO)/86.5\% (WESAD) Acc (2-class) \\
16.\cite{GSRBasedGeneric_PureEDA_ML_2023}& 2023 & Train: WESAD, S-Test, AffectiveROAD, and DS3; external test: UnseenTSST and UBFC-Phys & Manual artifact rejection, median filter, z-score (no-stress mean/std),resample to 4~Hz, segment selection, 4~Hz LPF, 30~s window & 47 (Multi-domain) (MIC, 40) & \textbf{Voting Ensemble} (GBT, RF, AdaBoost) & 90/10 subject split; also test on unseen datasets & 89.1\% (combined-data test) / 84.6\% (UnseenTSST) / 92.8\% (UBFC-Phys) Acc (2-class) \\

17.\cite{DeepLearningApproach_PureEDA_ML_2025}& 2025 & VerBIO, SWELL, WESAD, Neuro, Nurse Stress Prediction, RWDriving & Windowing, normalization, filtering, cvxEDA decomp. & 7 (Time) (None) & FNN, CNN, Wide and Deep Model & 5-fold CV, $\sim$10\% of subjects held out per fold & 88.7--98.1\% Acc (2-class) \\

18.\cite{20and32_HybridFeature_PureEDA_ML_2026}& 2026 & WESAD (chest) & Butterworth 0.05--5~Hz band-pass, 700$\to$20~Hz downsample, two-stage artifact correction, cvxEDA decomp. (20~Hz-adapted), 60~s window (50\% overlap) & 52 (Time, Supervised-AE latent) (None; AE bottleneck 64$\to$32) & \textbf{KNN (k=7)}, RF, XGB, SVM, NB, Ensembles & LOSO$^{\ddagger}$ & 98.62\% Acc (2-class) \\
\multicolumn{8}{l}{\textit{\textbf{Part II: Deep Learning (DL) Approaches}}} \\
\midrule
19.\cite{DNN_PureEDA_DL_2023}& 2023 & Self-collected; physical activity & 2~min window (1~min overlap) & \multicolumn{2}{p{0.40\textwidth}}{E2E; FCN} & Train on synthetic, test on real & 96\% (Synthetic) / 84\% (Real) Acc (2-class) \\
20.\cite{TinyStressNet_PureEDA_DL_2024}& 2024 & WESAD, S-Test, AffectiveROAD, DS3 & 4~Hz resample, band-pass filter, 30~s window & \multicolumn{2}{p{0.40\textwidth}}{E2E; \textbf{NAS} for TinyML} & 90/10 subject split & 85.98\% Acc (2-class), 49.4 kB / 3682 params vs. 89.1\% at 1.3 MB \\

21.\cite{INT8TinyML_EWCLite_PureEDA_DL_2026}& 2026 & WESAD&Interpolation, median filter, overlapping 4~s window, z-score norm& \multicolumn{2}{p{0.40\textwidth}}{E2E; TinyML-Compatible 1D-CNN with Personalization and quantization}& LOSO$^{\ddagger}$ & 65.90\% (global LOSO, INT8) / 66.35\% (float32) $\to$ 69.30\% after personalization (3-class)\\

22.\cite{FollowingResearchofDSVM_DL_2026}& 2026 &Self-collected; calm/stress movies  & 4~Hz LPF, Gaussian smoothing, CDA (Ledalab) decomp., derivative, 1--40~s window, z-score & \multicolumn{2}{p{0.40\textwidth}}{E2E; 1D-CNN, Temporal Convolutional Networks, InceptionTime, Time-Series Transformer, \textbf{PatchTST}}& LOSO & $0.863 \pm 0.053$ Acc, $0.852 \pm 0.056$ F1 (2-class) \\

\bottomrule
\end{tabular}
\end{adjustbox}
\caption{Summary of unimodal studies using only EDA for stress detection. The table is divided into Part I (ML) and Part II (DL). The best-performing model in each study is indicated in \textbf{bold}.
\textbf{Features Column Explanation}: Data are presented as follows: \textit{Total Number of Extracted Features (Feature Domains) (Feature Selection Method, Final Selected Count)}.
\textbf{Validation Notes}:
$^{\dagger}$ The evaluation partition is not subject-independent: records from the same subject appear in both the training and test partitions. 
$^{\ddagger}$ The partition is subject-independent, but a quantity used by the final model (a scaler, a projection, a feature subset, or a feature count) takes a value that depends on data from the test partition.
\textbf{Abbreviations}:
\textit{Models}: Autoencoder (AE), Convolutional Neural Network (CNN), Decision Tree (DT), Extreme Learning Machine (ELM), Fully Convolutional Network (FCN), Feedforward Neural Network (FNN), Gradient Boosted Trees (GBT), $k$-Nearest Neighbors (KNN), Linear Discriminant Analysis (LDA), Logistic Regression (LR), Neural Architecture Search (NAS), Naïve Bayes (NB), Probabilistic Neural Network (PNN), Quadratic Discriminant Analysis (QDA), Random Forest (RF), Stacking Ensemble Learning (SEL), Support Vector Machine (SVM).
\textit{Validation}: Cross-Validation (CV), Leave-One-Out Cross-Validation (LOOCV), Leave-One-Subject-Out (LOSO).
\textit{Others}: Maximal Information Coefficient (MIC), Trier Social Stress Test (TSST).}
\label{tab_unimodal_eda_stress_detection_studies}
\end{sidewaystable*}
}

{\color{\revisioncolor}
\subsection{Unimodal EDA Methods}
\label{sec_unimodal_methods}
Table~\ref{tab_unimodal_eda_stress_detection_studies} summarizes 22 unimodal EDA studies. Beyond the hardware convergence documented in Section~\ref{sec_background_and_related_work}, three aspects of research practice have converged: recent studies predominantly use public datasets, particularly WESAD; 8/9 studies published since 2022 that apply decomposition use cvxEDA; and subject-independent partitioning has become standard practice. Reported performance has not converged due to a lack of comparability across studies. On WESAD, accuracy ranges from 65.9\%~\cite{INT8TinyML_EWCLite_PureEDA_DL_2026} to 98.6\%~\cite{20and32_HybridFeature_PureEDA_ML_2026}. Both~\cite{LowCost_PureEDA_ML_2022} and~\cite{FrequencySpectrumAnalysis_PureEDA_ML_2023} reported binary-classification accuracy on WESAD under LOSO, achieving 97.03\% and 84.87\%, yet the two studies also differed in their features, classifiers, window lengths, step sizes, and normalization procedures, so the 12-percentage-point gap cannot be attributed to any single factor. The few studies that partly share a protocol come from the same research group~\cite{FeasibilityStudy_PureEDA_ML_2022, EDAisBest_PureEDA_ML_2022, ZhuLiLi_JBHI_PureEDA_ML_2023}. Each row of Table~\ref{tab_unimodal_eda_stress_detection_studies} therefore reports what one particular method achieved on one particular dataset under one particular protocol, not what the method itself can achieve.

Four protocol choices account for much of this spread:
\begin{itemize}
    \item \textbf{Subject separation} has converged: every study that trains and tests on the same participants ($\dagger$ in Table~\ref{tab_unimodal_eda_stress_detection_studies}) was published in 2022 or earlier. One study trained and tested within each subject~\cite{automaticworkloadestimation_PureEDA_ML_2016}; five partitioned folds by recording without separating participants~\cite{physicalandpsychologicalstressorsn_PureEDA_ML_2017, RealDriving_PureEDA_ML_2018, OpenSourceTool_PureEDA_ML_2020, electronicnose_PureEDA_ML_2020, CDACVX_DecompoistionComparision_2022}; and one used a random 70/15/15 split~\cite{DSVM_PureEDA_ML_2020}. The same research group later reevaluated that dataset with a Transformer under LOSO and obtained an F1 score of 0.85~\cite{FollowingResearchofDSVM_DL_2026}, below the 0.92 reported in 2020.
    \item \textbf{Filtering} remains inconsistent. Eight studies specify a low-pass filter with a numerical cutoff~\cite{physicalandpsychologicalstressorsn_PureEDA_ML_2017, XGBOOST_PureEDA_ML_2019, PRESURGERY_PureEDA_ML_2020, OpenSourceTool_PureEDA_ML_2020, DSVM_PureEDA_ML_2020, LowCost_PureEDA_ML_2022, GSRBasedGeneric_PureEDA_ML_2023, FollowingResearchofDSVM_DL_2026}, and one applied a band-pass~\cite{20and32_HybridFeature_PureEDA_ML_2026}; six report filtering that is incompletely specified or of another kind, covering median filtering alone, a 50-Hz notch with moving-average smoothing, and unspecified smoothing~\cite{RealDriving_PureEDA_ML_2018, TinyStressNet_PureEDA_DL_2024, DeepLearningApproach_PureEDA_ML_2025, INT8TinyML_EWCLite_PureEDA_DL_2026, automaticworkloadestimation_PureEDA_ML_2016, CDACVX_DecompoistionComparision_2022}; and seven report no independent conventional filter~\cite{electronicnose_PureEDA_ML_2020, FeasibilityStudy_PureEDA_ML_2022, EDAisBest_PureEDA_ML_2022, AcuteStressStateClassification_PureEDA_ML_2023, FrequencySpectrumAnalysis_PureEDA_ML_2023, ZhuLiLi_JBHI_PureEDA_ML_2023, DNN_PureEDA_DL_2023}.
    \item \textbf{Segmentation} is equally inconsistent. Window lengths range from 3.5~s to 5~min, with 30~s and 60~s together covering 8/22 studies. In~\cite{XGBOOST_PureEDA_ML_2019}, increasing the step size on WESAD from 0.25\,s to 30\,s, with features and classifiers unchanged, improved the F1 score of every tested classifier by more than 11 percentage points; the authors attributed this to redundancy between adjacent windows, yet subsequent work continued to use a 0.25\,s step~\cite{LowCost_PureEDA_ML_2022}. Segmentation also fixes detection latency, which~\cite{FrequencySpectrumAnalysis_PureEDA_ML_2023} expressed as $\mathrm{TTD}=t_{\mathrm{length}}+(\omega-1)t_{\mathrm{step}}$, allowing the overlap to be tuned for the best performance under a fixed latency budget.
    \item \textbf{Information leakage} appears in four of the 22 studies. In~\cite{LowCost_PureEDA_ML_2022,PRESURGERY_PureEDA_ML_2020,AcuteStressStateClassification_PureEDA_ML_2023}, the test set served as the validation set, so features were selected on the data used to report performance. 
    In~\cite{electronicnose_PureEDA_ML_2020}, a supervised LDA projection was fitted on all measurements before the subsequent five-fold cross-validation, allowing test-sample labels to influence the projection and potentially contributing to the reported 100\% accuracy.
    In~\cite{20and32_HybridFeature_PureEDA_ML_2026}, the feature scaler was fitted and the hyperparameter search conducted on the complete dataset before LOSO cross-validation.
\end{itemize}

Inter-subject variability affects accuracy once the protocol is fixed. With the data, features, and classifier held constant, subject-dependent and subject-independent accuracy differed by 21.6 percentage points on an emotion-recognition dataset~\cite{FeatureCombine_EDA_EmotionRecognition_2021}. Three approaches address this at different stages of the pipeline: Clustering subjects and training a separate classifier for each cluster improved three-class accuracy on held-out subjects by 6.1 percentage points~\cite{PRESURGERY_PureEDA_ML_2020}. Subject-wise feature normalization increased accuracy from 85.3\% to 94.7\%~\cite{LowCost_PureEDA_ML_2022}. After deployment, FT on the target user with 40 adaptation windows improved accuracy on WESAD by 3.4 percentage points~\cite{INT8TinyML_EWCLite_PureEDA_DL_2026}.

The dataset itself accounts for much of the reported number. With features and classifier held constant, one pipeline achieved 66.3\% on Neuro and 92.9\% on VerBIO~\cite{ZhuLiLi_JBHI_PureEDA_ML_2023}. Nearly all studies nonetheless estimate performance only within the dataset used for training; \cite{GSRBasedGeneric_PureEDA_ML_2023} is an exception, training on four datasets combined and reaching 84.6\% on a dataset held out entirely. 
The 22 studies also draw on only 9 of the 26 datasets cataloged in Table~\ref{tab_public_eda_stress_datasets}, leaving most of the cataloged scenarios, and the real-world conditions they stand in for, untested.

Two lines of work have tried to move beyond standard time- and frequency-domain statistics. The first learns representations from the signal using DL, and its returns have so far been limited. The NAS-derived model in~\cite{TinyStressNet_PureEDA_DL_2024} reached 85.98\% accuracy at 1/26 the size of the handcrafted-feature model it was benchmarked against, which reached 89.1\%~\cite{GSRBasedGeneric_PureEDA_ML_2023}; its advantage lay in compactness, not accuracy. In~\cite{20and32_HybridFeature_PureEDA_ML_2026}, an autoencoder whose latent space was supervised by stress labels achieved a LOSO accuracy of 89.0\%, more than 6 percentage points below the 95.1\% obtained from 20 statistical features under the same data, protocol, and classifier. The second designs representations by hand from EDA physiology: band-limited sympathetic-power measures and their time-varying refinement~\cite{EDASymp_NonlinearMethod_2016,TVSymp_Decomposition_2016}, lightweight regression-based slope and residual-area features~\cite{2feature_NonlinearMethod_2019}, and nonlinear phase-space complexity measures~\cite{ComEDA_NonlinearMethod_2022,MComEDA_NonlinearMethod_2024}.}

\label{sec_unimodal_methods_end}
\subsection{Insights from Multimodal Methods}
\label{sec_insights_from_multimodal_methods}
This paper focuses on unimodal EDA, so multimodal work is reviewed here for design principles that carry over. The bridge is decomposition: separating EDA into tonic and phasic components turns one signal into several correlated channels, so methods built for combining modalities apply directly to combining these channels.
\subsubsection{Architectural Evolution}
\label{sec_architectural_evolution}
Multimodal work has largely followed a feature extraction–fusion–classification pipeline. It depends on manual feature engineering and usually feeds EDA in whole without component separation \cite{EDAPPGACCST_Feature_ML_MultiModal_2020,contextbasedstressdetector_ML_MultiModal_2017,PPGEDAST_ledalab_ML_MultiModal_2017,EDAPPG_ACCST_cvxEDA_ML_MultiModal_2019,WESADComparision_ML_MultiModal_2020}. 
Hybrid designs that input handcrafted features into a neural network keep that engineering burden and add computational cost and opacity \cite{NNComparison_DL_MultiModal_2013,ECGEEGRESP_DL_MultiModal_2019,EDABVP_Feature_window_DL_MultiModal_2020,WESADTOUX_DL_MultiModal_2021,BVPEDA_Feature_DL_MultiModal_2022}. E2E designs that directly take raw or lightly processed signals as input can automatically learn signal representations, but lack interpretability \cite{EDAPPGST_DL_MultiModal_2020,MLPCNN_ETE_DL_MultiModal_2020}. Where fusion happens matters: \cite{FusionComparsion_DL_MultiModal_2021} reports early feature interaction outperforms late decision-level fusion. This finding motivates cross-modal calibration modules at high-level feature layers \cite{MMTM_DL_MultiModal_2021}, hierarchical architectures with intermediate fusion \cite{HierarchicalCNN_DL_MultiModal_2023,HierarchicalDNN_DL_MultiModal_2021}, and dual-channel frameworks that separate modality-invariant from modality-specific features \cite{crossdomain_DL_MultiModal_2022}. The dual-channel design is consistent with the finding that signals driven by the autonomic nervous system transfer across modalities far better than motion signals \cite{CrossModalInvestigation_DL_MultiModal_2025}. Architectures borrowed from computer vision, including ResNeXt-style parallel convolutions \cite{StressNeXt_DL_MultiModal_2022}, CNN–Transformer hybrids \cite{CNNTransformer_DL_MultiModal_2023}, and FCN–LSTM stacks \cite{FCNLSTM_DL_MultiModal_2024}, trade local feature extraction against global temporal modeling.
\subsubsection{Signal Representation}
\label{sec_signal_representation}
A second line of work converts 1D time series into 2D images so that image models can be applied. Reported encodings include continuous recurrence plots \cite{CONTPR_DL_MultiModal_2021}; stacked feature maps from the FFT, cube-root transforms and constant-Q transforms \cite{FFTCubeRootCQT__DL_MultiModal_2022}; and grayscale spectrograms \cite{RawSignal_To_GrayscaleSpectrogram_2024}. Gramian angular summation and difference fields are used more often than Markov transition fields because they cost less to compute; their reported gains over SVMs are attributed to preserving temporal dependencies in pixel space \cite{GAFs_DL_MultiModal_2024,SignalToImage_DL_MultiModal_2025}. Label-free alternatives have also been explored, including autoencoders with pseudo-inverse learning and hierarchical frequency-domain representations \cite{EnsembleClassificationModel_DL_MultiModal_2021,HierarchicalAutoencoder_DL_MultiModal_2023}.
\subsubsection{Context, Data, and Prediction Targets}
\label{sec_context_data_and_prediction_targets}
Moving from the laboratory to everyday settings makes physical activity hard to separate from psychological stress. Context-based frameworks use auxiliary sensors, typically accelerometers, to tell the two apart \cite{contextbasedstressdetector_ML_MultiModal_2016,contextbasedstressdetector_ML_MultiModal_2017,EDAPPG_ACCST_cvxEDA_ML_MultiModal_2019}, and adaptive versions use motion to switch between context-specific classifier branches \cite{Improvedmethod_ML_MultiModal_2024}. Two pitfalls share the same root: reliance on information available only in experimental settings. Experimental phase labels are unavailable in everyday life \cite{EDABVP_Context_ML_MultiModal_2021}, while continuous self-reports are unrealistic outside controlled studies \cite{Survey_Personal_DL_MultiModal_2021}.

Whether laboratory or in-the-wild data is preferable depends on the learning regime. Under supervised learning with limited data, clean laboratory recordings might generalize to noisy settings better than data collected in the wild \cite{EDAHRV_labToWild_ML_MultiModal_2020,labtoEnv_ML_MultiModal_2025}. Under SSL, the scale and diversity of the pretraining corpus matter more, and large unconstrained daily-life datasets outperform laboratory data as pretraining sources \cite{CrossDataset_Generalizability_SSL_Multimodal_2026}. Scenarios have widened accordingly, covering programming competitions, public speaking, speech, simulated threats, and specific populations such as individuals with mild cognitive impairment \cite{EDAPPG_ACCST_cvxEDA_ML_MultiModal_2019,publicspeaking_ML_MultiModal_2021,EDAandSpeech_ML_MultiModal_2021,Airraidsiren_ML_MultiModal_2024,ECGEDA_MCI_ML_MultiModal_2020}. Ensembles trained on merged datasets and pretraining-plus-FT are the two common responses to limited data \cite{EnsembleDataset_ML_MultiModal_2023,PreDLmodel_DL_MultiModal_2023}. Prediction targets are shifting as well, from binary classification toward regression on continuous stress scores, which keeps the graded information that binary labels discard and supports proportionate intervention \cite{StressScore_ML_MultiModal_2020,Regression_Classification_ML_MultiModal_2020}.
{\color{\revisioncolor}
\section{Experiments}
\label{sec_experiments}
\begin{table}[t]
  \begin{center}
    \caption{The 116 extracted EDA features. AUP denotes the average area under the phasic curve, and PSD denotes power spectral density. Features are computed from clean EDA, tonic, and phasic signals, except for SCR Events and Sparse Driver.}
    \label{tab_extracted_eda_features}
    \scriptsize
    \setlength{\tabcolsep}{2pt}
    \begin{tabularx}{\columnwidth}{l|X}
      \hline
      \textbf{Feature Group (Num.)} & \textbf{Parameters}\\
      \hline
      Time: SCR Events (7) & \{num, mean,std,max\}Peak, avg\{Rise,Fall\}Time, AUP\\
      Time: Sparse Driver (4) & \{num,mean,std,max\}Driver \\
      Time: Statistics (12) & mean, std, skew, kurt\\
      Time: Hjorth (6) & mobility, complexity\\
      \hline
      Freq: Spectral Location (6) & centroid, dominant frequency\\
      Freq: Spectral Amplitude (9) & mean, skew, kurt\\
      Freq: PSD, Welch (12) & mean, std, skew, kurt\\
      \hline
      Time-Freq: Wavelet (45) & db4, 4 levels: mean, std, entropy $\times$ \{cA4, cD4--cD1\}\\
      \hline
      Entropy (15) & sample, permutation, fuzzy, spectral, multiscale\\
      \hline
    \end{tabularx}
  \end{center}
\end{table}

The preceding review shows that reported results are difficult to compare. We therefore establish a unified subject-independent benchmark on five datasets, in which each experiment fixes one factor for the next. Section~\ref{sec_preprocessing_and_input_comparison} determines the preprocessing pipeline and DL input. Section~\ref{sec_model_comparison} then identifies the best overall model, RF, and the best DL model, BiLSTM, which are carried forward to the transfer analysis in Section~\ref{sec_transfer_analysis}.
\subsection{Experimental Setup}
\label{sec_experimental_setup}
\textbf{Datasets.} Neuro (subjects=20; 702/789 stress/non-stress segments), WESAD (15; 326/1147), MAUS (22; 880/440), AffectiveROAD (ROAD; 9; 1823/1494), and EmpaticaE4Stress (E4D; 29; 1396/754), comprising 95 subjects and 9751 segments in total, were selected to collectively span physical, cognitive, emotional, driving, and occupational stressors, providing the domain diversity required to assess generalizability.

\textbf{Preprocessing.} Signals were resampled to 4~Hz, $z$-scored per record, and split by label block into non-overlapping 30~s (120-sample) windows. Six branches were generated by combining two filtering conditions, no filtering and the 1~Hz second-order zero-phase Butterworth LPF adopted in prior EDA studies \cite{TaylorAutomatic_ArtifactRemoval_2015,IndicesofPain_DecompoistionComparision_2021}, with cvxEDA, sparsEDA, and ospEDA. We denote the decomposition input as $\mathrm{Raw}=y=p+t+\varepsilon$ and the returned Clean channel as $\mathrm{Clean}=p+t$, where $p$, $t$, and $\varepsilon$ denote the phasic, tonic, and residual components, respectively. Each algorithm uses its authors' published default settings.

\textbf{Inputs.} ML uses the 116 features listed in Table~\ref{tab_extracted_eda_features} after features correlated above $|r|=0.95$ are removed (leaving 84 for the selected preprocessing pipeline); DL uses the two-channel input $[\mathrm{Clean},\,t]$, and models with configurable depth default to three layers.

\textbf{Protocol.} Within-dataset and pooled results use LOSO: \emph{macro} denotes LOSO within each dataset followed by an unweighted mean across the five datasets, while \emph{pooled} denotes one LOSO loop over the merged set of 95 subjects. Transfer instead holds out entire datasets, with training performed on the complete source datasets and testing performed on every subject in a disjoint target dataset. Six metrics (ACC, F1, sensitivity, specificity, Matthews correlation coefficient (MCC), and AUC) are computed for each held-out subject and averaged with equal weight assigned to each subject. Given the opposing class imbalances across datasets (WESAD vs.\ MAUS), we report MCC \cite{WhywechooseMCC} throughout. Unlike accuracy and F1, MCC is high only when both classes are classified correctly. We report the other metrics when they distinguish failure modes that MCC conflates. ML classifiers use their library default settings. DL training is fixed at 100 epochs with a batch size of 64 and AdamW ($10^{-3}$, weight decay $0.01$); FT runs for 30 epochs at $10^{-4}$. Neither ML nor DL uses hyperparameter tuning, class weights, or a validation split; DL uses neither early stopping nor a scheduler. All runs use a seed of 42 and fit scalers using only the training data. Paired subject-wise differences are tested using the Wilcoxon signed-rank test with Holm correction within each family.
\subsection{Preprocessing and Input Comparison}
\label{sec_preprocessing_and_input_comparison}
\begin{table}[h]
\centering
\caption{Preprocessing comparison.}
\label{tab_preprocessing_comparison}
\small
\resizebox{\columnwidth}{!}{
\begin{tabular}{l l c c c c c c}
\toprule
\multirow{2}{*}{Dataset} & \multirow{2}{*}{Classifier}
& \multicolumn{2}{c}{cvxEDA}
& \multicolumn{2}{c}{sparsEDA}
& \multicolumn{2}{c}{ospEDA} \\
\cmidrule(lr){3-4}
\cmidrule(lr){5-6}
\cmidrule(lr){7-8}
& & {No Filter} & {1~Hz LPF}
  & {No Filter} & {1~Hz LPF}
  & {No Filter} & {1~Hz LPF} \\
\midrule

\multirow{5}{*}{\shortstack{\textit{Macro}\\[2pt]\scriptsize(rank)}}
& SVM & $\mathbf{0.409}^{(1)}$ & $0.401^{(2)}$
      & $0.328^{(5)}$ & $0.305^{(6)}$
      & $0.338^{(4)}$ & $0.344^{(3)}$ \\
& RF  & $0.437^{(2)}$ & $\mathbf{0.444}^{(1)}$
      & $0.367^{(5)}$ & $0.364^{(6)}$
      & $0.395^{(3)}$ & $0.375^{(4)}$ \\
& MLP & $0.393^{(2)}$ & $\mathbf{0.406}^{(1)}$
      & $0.361^{(3)}$ & $0.355^{(4)}$
      & $0.248^{(5)}$ & $0.245^{(6)}$ \\
& Transformer & $\mathbf{0.389}^{(1)}$ & $0.379^{(2)}$
      & $0.353^{(4)}$ & $0.359^{(3)}$
      & $0.221^{(6)}$ & $0.228^{(5)}$ \\
\cmidrule(lr){2-8}
& All & $0.407^{(2)}$ & $\mathbf{0.408}^{(1)}$
      & $0.352^{(3)}$ & $0.346^{(4)}$
      & $0.301^{(5)}$ & $0.298^{(6)}$ \\

\midrule

\multirow{5}{*}{\shortstack{\textit{Pooled}\\[2pt]\scriptsize(rank)}}
& SVM & $\mathbf{0.319}^{(1)}$ & $0.314^{(2)}$
      & $0.198^{(5)}$ & $0.183^{(6)}$
      & $0.238^{(3)}$ & $0.238^{(4)}$ \\
& RF  & $0.321^{(2)}$ & $\mathbf{0.331}^{(1)}$
      & $0.242^{(5)}$ & $0.223^{(6)}$
      & $0.313^{(4)}$ & $0.315^{(3)}$ \\
& MLP & $0.281^{(2)}$ & $\mathbf{0.298}^{(1)}$
      & $0.181^{(5)}$ & $0.196^{(3)}$
      & $0.179^{(6)}$ & $0.192^{(4)}$ \\
& Transformer & $0.287^{(2)}$ & $\mathbf{0.307}^{(1)}$
      & $0.199^{(4)}$ & $0.223^{(3)}$
      & $0.105^{(6)}$ & $0.135^{(5)}$ \\
\cmidrule(lr){2-8}
& All & $0.302^{(2)}$ & $\mathbf{0.312}^{(1)}$
      & $0.205^{(6)}$ & $0.206^{(5)}$
      & $0.209^{(4)}$ & $0.220^{(3)}$ \\

\bottomrule
\end{tabular}
}
\end{table}
\textbf{Decomposition.} As shown in Table~\ref{tab_preprocessing_comparison}, the two cvxEDA conditions ranked top two in both macro MCC and pooled MCC for every model. Without filtering, the average macro MCC across the four models was $0.407$ for cvxEDA, compared with $0.352$ for sparsEDA and $0.301$ for ospEDA. Per-record $z$-score normalization removes the $\mu$S scale and produces negative values, so the structural assumptions of each decomposition algorithm determine whether its outputs remain informative. cvxEDA models the tonic component using an unconstrained B-spline basis, offset, and trend; its all-zero rate remains below $1\%$ for every output channel. ospEDA clips negative tonic estimates to zero and uses fixed thresholds originally defined in $\mu$S. It therefore returns an all-zero tonic component in $60.0\%$ of windows, and the max--min difference of its Clean channel is less than one-tenth of that of the Raw input in $50.2\%$ of windows. sparsEDA is designed to localize a small number of discrete sudomotor impulses and imposes stronger sparsity than cvxEDA. This property is useful for event localization, but suppressing weak or overlapping responses removes phasic dynamics used by our classifiers with $30$~s windows: its phasic component is identically zero in $23.4\%$ of windows, and its driver is identically zero in $54.0\%$. Our multi-dataset, subject-wise comparison consistently supports cvxEDA across four classifiers.

\textbf{Filtering.} Under cvxEDA, adding the filter changes macro MCC by only $0.001$ and pooled MCC by $0.010$; none of the eight comparisons across the four models is significant after Holm correction ($p=0.504$--$1.000$). We also compare each window before and after filtering using the relative difference
$
\delta(u,v)=\frac{\mathrm{RMS}(u-v)}{\mathrm{RMS}(u)},
$
where $u$ and $v$ are the unfiltered and filtered versions, respectively. The median difference is $0.69\%$ at the decomposition input but only $0.09\%$ between the two cvxEDA Clean outputs. This reduction is consistent with the cvxEDA model $y=p+t+\varepsilon$, where the noise component $\varepsilon$ is excluded from $\mathrm{Clean}=p+t$ and is therefore not passed to the classifier. We therefore use cvxEDA without the additional LPF.

\begin{table}[h]
\centering
\caption{Channel input comparison.}
\label{tab_channel_input_comparison}
\small
\setlength{\tabcolsep}{3.5pt}
\renewcommand{\arraystretch}{0.95}
\resizebox{\columnwidth}{!}{
\begin{tabular}{l l c c c c c c c}
\toprule
\multirow{2}{*}{Model} & \multirow{2}{*}{Dataset} & \multicolumn{3}{c}{Single Channel} & \multicolumn{4}{c}{Multi-channel} \\
\cmidrule(lr){3-5}
\cmidrule(lr){6-9}
& & C & T & P & C+T & C+P & T+P & C+T+P \\
\midrule
\multirow{2}{*}{Transformer} & \shortstack{\textit{Macro} \scriptsize(rank)} & $0.376^{(4)}$ & $0.365^{(6)}$ & $0.245^{(7)}$ & $\mathbf{0.389}^{(1)}$ & $0.380^{(3)}$ & $0.376^{(4)}$ & $0.382^{(2)}$ \\
& \shortstack{\textit{Pooled} \scriptsize(rank)} & $0.280^{(5)}$ & $0.210^{(6)}$ & $0.046^{(7)}$ & $0.287^{(4)}$ & $0.313^{(2)}$ & $0.305^{(3)}$ & $\mathbf{0.317}^{(1)}$ \\
\midrule
\multirow{2}{*}{MLP} & \shortstack{\textit{Macro} \scriptsize(rank)} & $0.367^{(5)}$ & $0.366^{(6)}$ & $0.274^{(7)}$ & $0.393^{(2)}$ & $0.393^{(2)}$ & $\mathbf{0.396}^{(1)}$ & $0.393^{(2)}$ \\
& \shortstack{\textit{Pooled} \scriptsize(rank)} & $0.251^{(6)}$ & $0.225^{(7)}$ & $0.266^{(5)}$ & $0.281^{(3)}$ & $0.292^{(2)}$ & $\mathbf{0.293}^{(1)}$ & $0.281^{(3)}$ \\
\bottomrule
\end{tabular}
}
\vspace{5pt}
{\scriptsize C: clean; T: tonic; P: phasic.}
\end{table}
As shown in Table~\ref{tab_channel_input_comparison}, we evaluated seven input-channel configurations. Multi-channel configurations ranked among the top four in all model--evaluation settings and generally outperformed single-channel inputs. This advantage may arise from representational redundancy: providing multiple related views of the same signal makes both its overall and decomposed structure more accessible to the model. Pairwise comparisons among the four multi-channel inputs revealed no significant differences after Holm correction (adjusted $p=0.053$--$1.000$), with no configuration showing a consistent advantage. This similarity is consistent with $\mathrm{Clean}=p+t$, because any two channels contain sufficient information to reconstruct the third. We therefore use $\mathrm{Clean}+t$ as a representative multi-channel configuration in the subsequent experiments; any of the four multi-channel configurations could equally serve as the fixed input based on these results.
\label{sec_preprocessing_and_input_comparison_end}
\subsection{Model Comparison}
\label{sec_model_comparison}
\begin{figure}[h]
    \centering
    \includegraphics[width=1\columnwidth]{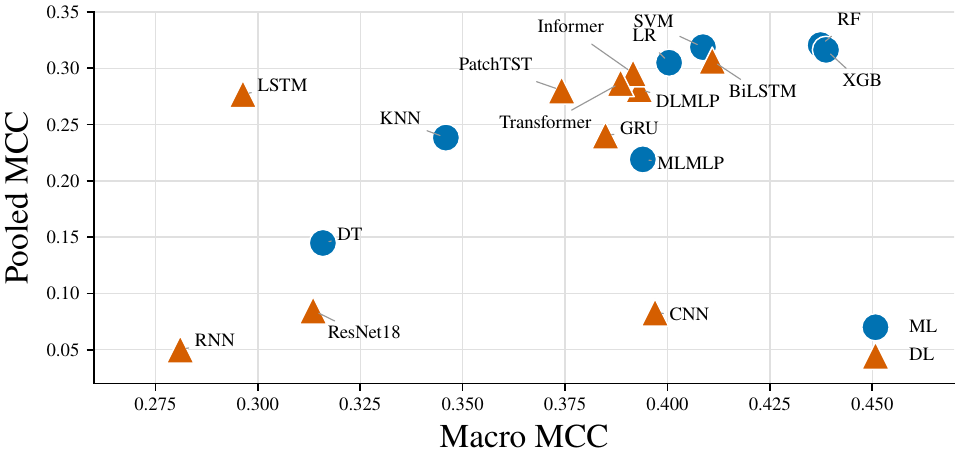}
    \caption{Model comparison.}
    \label{fig_model_comparison}
\end{figure}
As shown in Fig.~\ref{fig_model_comparison}, \textbf{ML models occupy the top positions in the ranking under both settings.} Under the small-sample LOSO setting, physiology-driven features already project the EDA signal onto a low-dimensional manifold aligned with the task structure. The ML models therefore need to learn only the decision boundary, whereas E2E DL models must also learn a signal representation from the raw data. BiLSTM is the best-performing DL model. RNN and unidirectional LSTM compress each window into a terminal hidden state, whereas BiLSTM retains information from both the early and late portions of the window, thereby reducing dependence on the position of the response within the window. Among the ML models, RF performs best overall, followed closely by XGB. Through multi-tree averaging and stage-wise correction, respectively, they eliminate unstable subject-specific rules retained by a single DT, preserving recurring patterns and resisting small-sample fluctuations.

\label{loc_model_failure_analysis}
\textbf{Two groups of DL models incorporate mechanisms that do not benefit this task}: CNN and ResNet18 perform moderately under macro evaluation ($0.397$, $0.314$) but collapse under pooled training ($0.083$, $0.085$; sensitivity $0.949$, $0.842$ versus specificity $0.109$, $0.239$), predicting stress for almost every sample. 
They are also the only two DL models that combine BatchNorm with global average pooling, suggesting that this combination may make them particularly sensitive to cross-subject distribution shifts. For a held-out test subject, mismatched BatchNorm statistics can systematically shift the normalized features, and global average pooling can carry this shift directly into the final classifier layer.
\label{loc_transformer_model_comparison}
The two time-series-specific Transformers, Informer and PatchTST, are indistinguishable from the standard Transformer under both settings (Holm-adjusted $p=0.603$--$0.983$); their sparse-attention and patching mechanisms target long sequences and have little to exploit in a 120-sample window.

\begin{figure}[htbp]
    \centering
    \includegraphics[width=1\columnwidth]{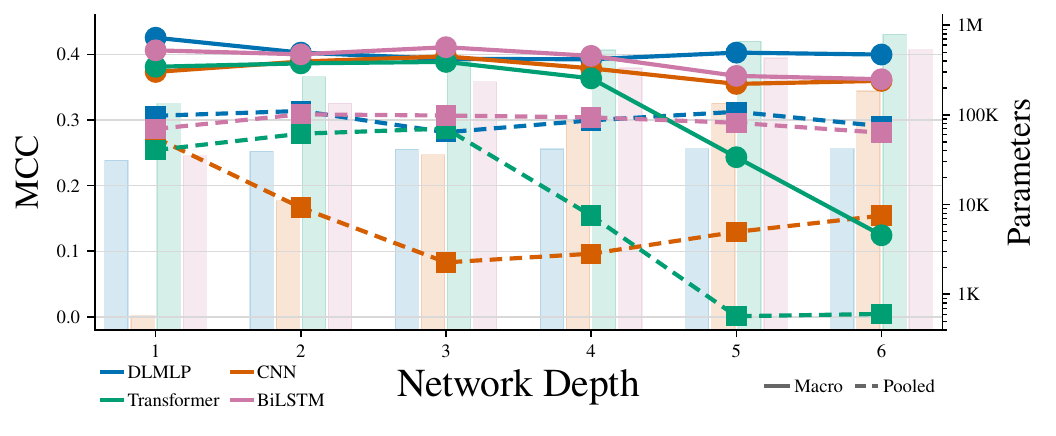}
    \caption{Depth comparison: MCC (lines, left axis) and trainable parameters (bars, right axis).}
    \label{fig_depth_comparison}
\end{figure}

Fig.~\ref{fig_depth_comparison} reports the parameter counts and performance of four representative models with one to six layers, while all other model and training settings are held fixed. The parameter count of DLMLP remains on the order of 40k, while that of BiLSTM grows from 35k to 0.5M; the performance of both models remains stable, indicating that a shallow, low-parameter architecture is sufficient to extract the informative content of EDA. In contrast, CNN and Transformer degrade as depth increases, collapsing at 2--4 and 5--6 layers, respectively, to nearly constant predictions of the stress class, with sensitivity approaching 1. Hence, the parameter growth induced by additional depth does not guarantee improved performance and may amplify architecture-specific instability. Across all tested depths, the highest DL macro MCC is \(0.425\) from the one-layer DLMLP, still below RF (\(0.437\)) and XGB (\(0.439\)), further confirming the performance advantage of ML.
\label{sec_model_comparison_end}

\subsection{Transfer Analysis}
\label{sec_transfer_analysis}
We use each dataset as the target in turn and the remaining datasets as sources. We evaluate four settings: training on one source, training on two sources, training on all four sources, and FT with labeled data from the target. In the comparison, ``Best observed 1 source'' and ``Best observed 2 sources'' denote the source or source pair with the highest MCC for each target and model. Four-source training uses three strategies. Pooled training combines all source samples, balanced training equalizes the different source dataset sizes, and the source ensemble trains one model per source and averages their prediction probabilities. FT updates the four-source pooled BiLSTM using the other subjects from the target dataset and tests it on the held-out target subject.

The ML model remains more stable than the DL model across the transfer experiments. RF exceeds BiLSTM by $0.077$, $0.055$, $0.151$, and $0.127$ macro MCC in single-source transfer, two-source transfer, balanced training, and the source ensemble, respectively. The two models perform similarly under pooled training ($p=0.778$). FT is evaluated only for BiLSTM.

\begin{figure}[h]
    \centering
    \setlength{\subfigtopskip}{0pt}
    \setlength{\subfigbottomskip}{0pt}
    \subfigure[RF]{%
        \includegraphics[width=0.49\columnwidth]{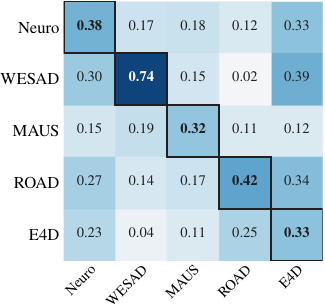}}%
    \hfill
    \subfigure[BiLSTM]{%
        \includegraphics[width=0.49\columnwidth]{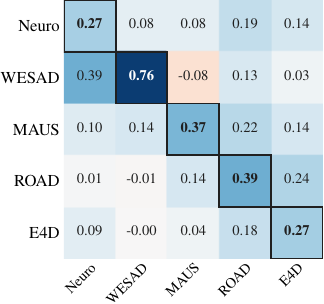}}
    \vspace{0pt}                
    \caption{Single-source transfer MCC. Rows are targets, columns sources; the boxed diagonal is within-dataset LOSO.}
    \label{fig_single_source_transfer_mcc}
    \vspace{-8pt}                
\end{figure}
\paragraph{Single-source transfer.}
All single-source transfer results are worse than within-dataset LOSO (Fig.~\ref{fig_single_source_transfer_mcc}), showing that performance depends strongly on dataset-specific acquisition settings and elicitation tasks.
ROAD and E4D are each other's best sources for both RF and BiLSTM. Despite their different elicitation paradigms (naturalistic driving vs.\ simulated office work), both contain continuous recordings of everyday-like activities, as opposed to recordings collected under tightly controlled laboratory protocols, which accounts for their transfer advantage and cross-model stability.
MAUS$\rightarrow$WESAD achieves a high AUC of $0.830$, but its sensitivity is $0.952$ and its specificity is only $0.034$. The model ranks the samples reasonably well but classifies almost all of them as stress, suggesting that adjustment of the decision threshold might correct much of the error.
\begin{table}[htbp]
\centering
\caption{Two-source transfer MCC, ranked per target. Each cell reports sources and MCC. N: Neuro; W: WESAD; M: MAUS; R: ROAD; E: E4D.}
\label{tab_two_source_transfer_horizontal}
\setlength{\tabcolsep}{2pt}
\renewcommand{\arraystretch}{0.8}
\resizebox{\columnwidth}{!}{
\begin{tabular}{l c c c c c c c c c}
\toprule
Target & Model & LOSO & Best 1 & Rank 1 & Rank 2 & Rank 3 & Rank 4 & Rank 5 & Rank 6 \\
\midrule
\multirow{2}{*}{Neuro}
& \raisebox{1.2ex}[0pt][0pt]{RF}     & \raisebox{1.2ex}[0pt][0pt]{0.38} & \shortstack{E\\0.33} & \shortstack{W+E\\0.27} & \shortstack{W+R\\0.22} & \shortstack{M+E\\0.22} & \shortstack{R+E\\0.18} & \shortstack{W+M\\0.11} & \shortstack{M+R\\0.09} \\
& \raisebox{1.2ex}[0pt][0pt]{BiLSTM} & \raisebox{1.2ex}[0pt][0pt]{0.27} & \shortstack{R\\0.19} & \shortstack{W+E\\0.19} & \shortstack{W+R\\0.18} & \shortstack{M+R\\0.13} & \shortstack{M+E\\0.12} & \shortstack{R+E\\0.11} & \shortstack{W+M\\0.09} \\
\midrule
\multirow{2}{*}{WESAD}
& \raisebox{1.2ex}[0pt][0pt]{RF}     & \raisebox{1.2ex}[0pt][0pt]{0.74} & \shortstack{E\\0.39} & \shortstack{N+E\\0.30} & \shortstack{M+E\\0.29} & \shortstack{N+M\\0.27} & \shortstack{N+R\\0.15} & \shortstack{R+E\\0.06} & \shortstack{M+R\\$-0.02$} \\
& \raisebox{1.2ex}[0pt][0pt]{BiLSTM} & \raisebox{1.2ex}[0pt][0pt]{0.76} & \shortstack{N\\0.39} & \shortstack{N+E\\0.30} & \shortstack{N+M\\0.23} & \shortstack{N+R\\0.18} & \shortstack{M+R\\0.13} & \shortstack{R+E\\$-0.12$} & \shortstack{M+E\\$-0.49$} \\
\midrule
\multirow{2}{*}{MAUS}
& \raisebox{1.2ex}[0pt][0pt]{RF}     & \raisebox{1.2ex}[0pt][0pt]{0.32} & \shortstack{W\\0.19} & \shortstack{N+W\\0.17} & \shortstack{W+R\\0.17} & \shortstack{N+R\\0.15} & \shortstack{N+E\\0.14} & \shortstack{R+E\\0.12} & \shortstack{W+E\\0.11} \\
& \raisebox{1.2ex}[0pt][0pt]{BiLSTM} & \raisebox{1.2ex}[0pt][0pt]{0.37} & \shortstack{R\\0.22} & \shortstack{R+E\\0.24} & \shortstack{W+R\\0.23} & \shortstack{N+R\\0.23} & \shortstack{N+E\\0.20} & \shortstack{W+E\\0.19} & \shortstack{N+W\\0.13} \\
\midrule
\multirow{2}{*}{ROAD}
& \raisebox{1.2ex}[0pt][0pt]{RF}     & \raisebox{1.2ex}[0pt][0pt]{0.42} & \shortstack{E\\0.34} & \shortstack{W+E\\0.37} & \shortstack{N+W\\0.33} & \shortstack{N+E\\0.31} & \shortstack{M+E\\0.25} & \shortstack{N+M\\0.25} & \shortstack{W+M\\0.21} \\
& \raisebox{1.2ex}[0pt][0pt]{BiLSTM} & \raisebox{1.2ex}[0pt][0pt]{0.39} & \shortstack{E\\0.24} & \shortstack{W+E\\0.30} & \shortstack{N+E\\0.28} & \shortstack{M+E\\0.21} & \shortstack{N+M\\0.17} & \shortstack{W+M\\0.07} & \shortstack{N+W\\0.06} \\
\midrule
\multirow{2}{*}{E4D}
& \raisebox{1.2ex}[0pt][0pt]{RF}     & \raisebox{1.2ex}[0pt][0pt]{0.33} & \shortstack{R\\0.25} & \shortstack{N+R\\0.28} & \shortstack{W+R\\0.21} & \shortstack{M+R\\0.20} & \shortstack{N+M\\0.19} & \shortstack{N+W\\0.17} & \shortstack{W+M\\0.02} \\
& \raisebox{1.2ex}[0pt][0pt]{BiLSTM} & \raisebox{1.2ex}[0pt][0pt]{0.27} & \shortstack{R\\0.18} & \shortstack{N+R\\0.24} & \shortstack{W+R\\0.18} & \shortstack{M+R\\0.17} & \shortstack{N+M\\0.13} & \shortstack{N+W\\0.07} & \shortstack{W+M\\$-0.01$} \\
\bottomrule
\end{tabular}
}
\end{table}

\paragraph{Two-source transfer.} Adding a second source does not consistently improve performance (Table~\ref{tab_two_source_transfer_horizontal}). Even the best two-source configuration is worse than the best single source for three RF targets and two BiLSTM targets. For WESAD as the target, E4D is the best individual source for RF and Neuro is the best for BiLSTM, yet combining E4D and Neuro reduces MCC by $0.09$ for both models. Conversely, for BiLSTM with E4D as the target, adding Neuro to ROAD increases sensitivity, specificity, AUC and MCC by $0.03$, $0.02$, $0.04$, and $0.06$, respectively. When the sources are well matched, increasing the amount of source data can improve both threshold-independent ranking ability and threshold-dependent classification performance.
\vspace{-10pt}
\begin{figure}[htbp]
    \centering
    \includegraphics[width=1\columnwidth]{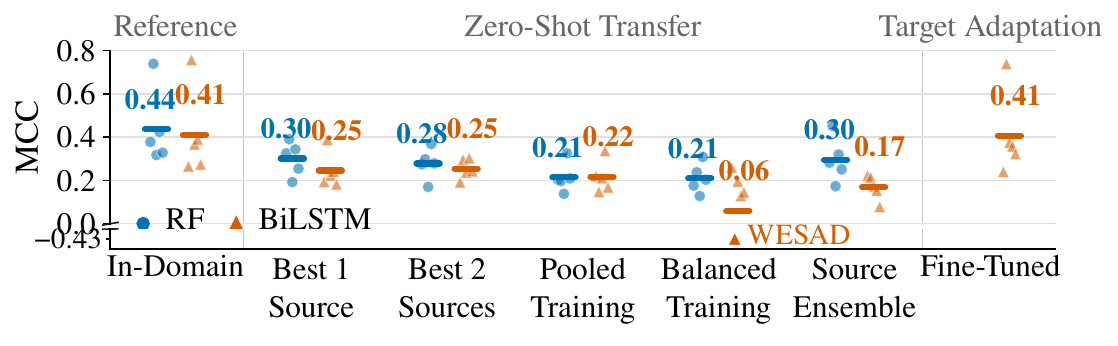}
     \vspace{-27pt}
    \caption{Comparison of the in-domain reference, zero-shot transfer, and target FT by MCC. Circles denote RF, triangles denote BiLSTM, markers denote the five target datasets, and horizontal bars and labels show their mean. FT is evaluated only for BiLSTM.}
    \label{fig_in_domain_transfer_and_fine_tuning_comparison}
     \vspace{-10pt}
\end{figure}
\paragraph{Four-source aggregation and adaptation.}
Pooled training performs worse than the best single-source and two-source configurations but better than their weakest configurations (Fig.~\ref{fig_in_domain_transfer_and_fine_tuning_comparison}). The diversity of the source data helps prevent performance collapse caused by source–target mismatch, but it still cannot match the best-matched source.
The gap between pooled training and within-dataset LOSO varies sharply across targets. WESAD falls from in-domain MCC values of $0.739$ and $0.757$ to $0.197$ and $0.148$ for RF and BiLSTM, respectively, representing the largest decreases among the five datasets. ROAD falls from $0.424$ and $0.390$ to $0.326$ and $0.337$; this decrease is the smallest among the five targets. Interestingly, the dataset that is easiest in-domain is the least stable under cross-domain transfer, whereas a dataset of moderate in-domain difficulty transfers most consistently.
For RF, balanced and pooled training perform similarly, whereas the source ensemble performs best. Training separate RFs preserves source-specific feature thresholds, while pooling forces a single set of trees to accommodate heterogeneous feature distributions, weakening their decision rules. BiLSTM shows the opposite pattern: pooled training performs best because the larger combined dataset enables it to learn temporal structures shared across sources. In contrast, the source ensemble learns four separate representations and combines only their predictions, limiting cross-source representation learning.
FT raises the macro-averaged MCC from $0.217$ to $0.407$, statistically indistinguishable from the within-domain value of $0.411$ ($p=0.915$). This recovery indicates that the source-trained model has learned stress-related patterns, but their direct application is limited by the distribution difference between the source and target datasets. FT adapts the pretrained model to the target distribution, allowing it to apply the learned classification capability effectively.

Taken together, the benchmark shows that reliable EDA-based stress detection depends more on consistent processing and evaluation protocols, effective signal representation, and domain compatibility than on model complexity or simply pooling more training data.
\label{sec_transfer_analysis_end}
\label{sec_experiments_end}
}

\section{Future Outlook and Discussion}
\label{sec_future_outlook_and_discussion}
\subsection{Robustness and Generalization}
\label{sec_robustness_and_generalization}
\subsubsection{Dynamic Temporal Adaptation}
\label{sec_dynamic_temporal_adaptation}
Fixing the analysis window imposes a trade-off between the information available to the model and the latency it can achieve. A single standard length limits flexibility, since effective detection may require different temporal resolutions depending on stressor dynamics. Longer windows theoretically carry more information, yet reported performance saturates as they grow, so window length must be chosen against a real-time latency budget \cite{EDABVP_Feature_window_DL_MultiModal_2020,FFTCubeRootCQT__DL_MultiModal_2022}. Future frameworks should therefore adjust the window dynamically: deep reinforcement learning agents, for instance, can optimize window length autonomously to balance prediction accuracy against response latency \cite{DynamicWindowSize_FutureDirection_2023}.
\subsubsection{Personalization and Domain Alignment}
\label{sec_personalization_and_domain_alignment}
Inter-subject physiological variability and scenario diversity remain major barriers to model generalization, motivating both rigorous evaluation across activities and populations \cite{CrossActivityandCrossPopulation_ML_MultiModal_2025,Reproducibility_ML_MultiModal_2020} and the development of personalized modeling approaches. Personalized modeling has responded with Multi-Task Learning \cite{MultitaskNN_DL_MultiModal_2017}, domain adaptation \cite{AdaptationandPersonalization_DL_MultiModal_2017}, demographic-aware feature engineering \cite{ECGEEGEDARSPEMG_PersonnalInfo_ML_MultiModal_2019,WESADTOUX_DL_MultiModal_2021}, and cluster-based models \cite{EDAECGRESEMG_Personnal_ML_MultiModal_2020}. Among these, unsupervised clustering is insufficient to capture inter-subject variability. Effective personalization calls for transfer learning and FT, with static descriptors (e.g., age) supplied to the model so that it adapts to individual response patterns \cite{SystematicEvaluationPersonalized_FutureDirection_2024,INT8TinyML_EWCLite_PureEDA_DL_2026}.

Furthermore, our experiments indicate that simple data aggregation may dilute informative signals, which makes advanced domain adaptation techniques necessary. Methods include Maximum Mean Discrepancy for aligning feature distributions \cite{MMD_FutureDirection_2017} and contrastive learning with related alignment algorithms for bringing different domains closer in the feature space \cite{CLDA_FutureDirection_2021}. Recent work that combines supervised contrastive learning with metadata shows promise for learning generalized stress representations that remain robust to individual differences \cite{contrastivelearning_Meta_FutureDirection_2025}. Domain generalization extends this approach further: modeling the decision boundary as an instance-conditioned distribution enables personalization for unseen users without target labels or retraining \cite{VIAStress_Personalization_2026}. In addition to signal distributions, stable psychological traits (e.g., stress mindset) can serve as modeling priors for personalization \cite{MindsetMatters_Personalization_2026}. The challenge also extends beyond individual physiological differences to the sensing hardware itself: ensuring reproducibility across heterogeneous consumer-grade devices remains a critical and often overlooked step toward widespread adoption \cite{ReproducibilityConsumerWearableSensors_FutureDirection_2025}.
\subsection{Data-Centric Paradigms}
\label{sec_data_centric_paradigms}
\subsubsection{Synthetic Data Augmentation}
\label{sec_synthetic_data_augmentation}
Although we have compiled 26 public datasets, their scale remains insufficient for training large foundation models. Beyond basic physiological modeling \cite{DNN_PureEDA_DL_2023}, more advanced generative models such as Generative Adversarial Networks (GANs), variational autoencoders, and diffusion models remain underexplored in this field \cite{DataGenerationSurvey_FutureDirection_2024}. For example, Conditional GANs with diversity constraints generate high-fidelity EDA signals while preventing mode collapse \cite{ConditionalGAN_FutureDirection_2022}.
\subsubsection{SSL and Foundation Models (FMs)}
\label{sec_self_supervised_learning_and_foundation_models}
Reliable ground truth remains elusive: activity labels are prone to confounding factors \cite{ActicityLabelisBad_FutureDirection_2025}, cortisol assays require obtrusive and delayed laboratory sampling that is unsuitable for continuous monitoring \cite{Cortisol_DL_MultiModal_2022}, and self-reports suffer from subjective bias. SSL offers a pathway to reduce dependence on labeled data by learning representations from unlabeled signals \cite{SenseandLearn_SSL_2021,CrossDataset_Generalizability_SSL_Multimodal_2026}. 
Although general sensor-based pretext tasks exist, signal-specific strategies, such as learning dynamic changes in physiological baselines through time-series prediction tasks, have proven more effective \cite{Islam_SSL_2023,Islam_OnlyEDA_SSL_2023}. Notably, tailoring data-augmentation strategies to EDA, including altering frequency components and distorting tonic and phasic components, within contrastive learning frameworks such as SimCLR has been shown to yield representations that outperform supervised baselines \cite{MIT_OnlyEDA_SSL_2023}. Building on this work, both EDA-specific FMs \cite{EDA_FoundationModel_SSL_2026} and multimodal wearable FMs \cite{MultivariateWearable_FoundationModel_SSL_2025} have recently emerged. However, the EDA-specific FM does not yet show a significant advantage over traditional ML approaches.
\subsection{Efficient and Trustworthy Deployment}
\label{sec_efficient_and_trustworthy_deployment}
\subsubsection{Edge Efficiency and Privacy}
\label{sec_edge_efficiency_and_privacy}
Deploying models on resource-constrained wearables requires rigorous optimization. TinyML quantization \cite{TinyML_lightEmployment_2025,INT8TinyML_EWCLite_PureEDA_DL_2026} and NAS (e.g., TinyStressNet and MC-QDSNN) can reduce computational overhead without compromising accuracy \cite{TinyStressNet_PureEDA_DL_2024,MCQDSNN_FutureDirection_2025}; MC-QDSNN further shows bio-inspired spiking neural networks offer a low-power alternative to traditional architectures. Privacy is the second constraint. Federated Learning (FL) trains models in a decentralized manner and preserves user privacy by keeping raw data local \cite{FederatedLearning1_FutureDirection_2023,FederatedLearning2_FutureDirection_2025}, at some cost in accuracy that must be managed \cite{FederatedLearning3_AccDecrease_FutureDirection_2022}. Pairing FL with SSL also removes the need for labels, which mitigates label leakage and improves performance across client distributions that are not independent and identically distributed \cite{FLwithSSL_FutureDirection_2026}.
\subsubsection{Explainability and Physiological Grounding}
\label{sec_explainability_and_physiological_grounding}
Integrating DL into clinical workflows requires models whose decisions can be inspected and traced to physiology. Post-hoc analysis using Integrated Gradients has confirmed that the rising and recovery phases of EDA peaks contribute most to stress prediction, in alignment with physiological theory \cite{IntegratedGradients_XAI_FutureDirection_2024}, while SHapley Additive exPlanations (SHAP) values validate feature importance in traditional models \cite{SHAP_XAI_FutureDirection_2024}.
\label{loc_physics_informed_eda}
{\color{\revisioncolor}Moving forward, Physics-Informed Neural Networks (PINNs), which embed governing equations into the training loss, provide a direct route toward physiologically grounded EDA models. PINNs have already incorporated local smoothness, Windkessel, and Navier--Stokes constraints for cardiovascular reconstruction \cite{Taylorconstraint_PINN_2023,Windkesselcardiovascularmodel_PINN_2026,WNavierStokesequations_PINN_2026}, as well as latent neural dynamics for EEG decoding \cite{EEG_PINN_2025}. Only one attempt has been made for EDA, constraining a first-order equation using questionnaire scores collected once per condition \cite{MultiTaskPINN_FutureDirection_2025}. Because these scores remain constant within each condition and have no mechanistic connection to sweat-gland dynamics, the resulting term is a generic regularizer and does not provide a physiological model of EDA.

A stronger direction is to derive constraints from established models of sweat-gland and sudomotor dynamics. cvxEDA \cite{cvxEDA_Decomposition_2016} could provide one concrete starting point: its Bateman convolution can be rewritten as a second-order differential equation, whose residual could then enter the PINN loss. Such a formulation would force the learned phasic response to remain consistent with a sparse, non-negative sudomotor driver, while a spline prior would keep the tonic component slow. More generally, such a formulation would change the nature of the decomposition itself. cvxEDA treats decomposition as a convex optimization problem that must be solved anew for every recording, whereas a PINN would treat it as a learning problem, so that a trained network could decompose a new signal in a single forward pass and could learn its physiological representation jointly with the stress classifier. The reconstruction and dynamics losses are computed from the signal alone and require no stress labels, so they would act as SSL objectives: unlabeled EDA could train the physiological representation, while labeled samples would contribute only through the task loss. Subject-specific physiological parameters could further adapt the model to individual sweat-gland dynamics while preserving its mechanistic structure.
}
\label{sec_efficient_and_trustworthy_deployment_end}
\section{Conclusion}
\label{sec_conclusion}
{\color{\revisioncolor}This study addresses the unvalidated conventions and fragmented practices in EDA-based stress detection by presenting the first survey to combine a review of the unimodal EDA pipeline with a benchmark. We compiled a resource of 26 public datasets, categorized 22 representative unimodal studies, and showed that the accuracies reported across these studies are not comparable.
Our benchmark therefore provides the first multi-dataset validation of cvxEDA and shows that feature-based ML outperforms E2E DL. The results identify methodological consistency, effective signal representation, and the ecological validity of data as the foundations of reliable stress detection, with limited returns from greater architectural complexity or simple data aggregation.
Building on these findings, we outline future directions across three dimensions: robustness and generalization under domain shift and inter-subject variability, data-centric paradigms such as generative augmentation and SSL to alleviate label scarcity, and efficient and trustworthy deployment on wearable devices. We believe this work will shape the next stage of EDA-based stress detection and accelerate its transition into reliable and deployable real-world systems.}
\label{sec_conclusion_end}
\bibliographystyle{IEEEtran}
\bibliography{main}
\end{document}